\documentclass[a4paper,11pt]{article}
\usepackage[utf8]{inputenc}

\usepackage[english]{babel}
\pdfoutput=1

\usepackage{graphicx,sidecap}
\usepackage{bm}
\usepackage{bbm}
\usepackage{braket}
\usepackage{amssymb}
\usepackage{cite}
\usepackage{mathrsfs}
\usepackage{amsmath}
\usepackage{xcolor}
\usepackage{hyperref}
\usepackage{enumerate}
\usepackage[toc,title]{appendix}
\usepackage{verbatim}
\usepackage[margin=30pt, bf, font=footnotesize, center, justification=justified]{caption}[2004/07/16]
\usepackage[T1]{fontenc} 
\usepackage{xcolor}
\usepackage{subfigure}
\usepackage{braket}
\usepackage[left=3cm,
            right=2.5cm,
            top=2.5cm,
            bottom=3cm
            ]{geometry}    
\hypersetup{colorlinks,bookmarksopen,bookmarksnumbered,
citecolor=red,
linkcolor=blue,
pdfstartview=green,
urlcolor=teal}

\date{}

\catcode`@=11 
\renewcommand\tableofcontents{%
  \section*{\contentsname}%
  \@starttoc{toc}%
}
\catcode`@=12

\begin{document} 

\title{\bf{Quench dynamics of R\'enyi negativities\\ and the quasiparticle picture}}
\maketitle

\vspace{-1.2cm}

\begin{center}
\author{Sara Murciano$^1$, Vincenzo Alba$^{2,3}$, and Pasquale Calabrese$^{1,4}$

\vspace{.5cm}

$^1${\small SISSA and INFN Sezione di Trieste, via Bonomea 265, 34136 Trieste, Italy}\\
$^2${\small Institute for Theoretical Physics, Universiteit van Amsterdam, Science Park 904, Postbus 94485, 1098 XH Amsterdam, The Netherlands}\\
$^3${\small Dipartimento di Ingegneria Industriale, Universit\`a degli Studi di Salerno, Via Giovanni Paolo II, 132 I-84084 Fisciano (SA), Italy}\\
$^3$ {\small International Centre for Theoretical Physics (ICTP), Strada Costiera 11, 34151 Trieste, Italy}\\
}
\end{center}

\vspace{.5cm}

\begin{abstract}
The study of the moments of the  partially transposed density matrix 
provides a new and effective way  of detecting bipartite entanglement in a many-body 
mixed state. This is valuable for cold-atom and ion-trap experiments, as well as in the general 
context of quantum simulation of many-body systems. In this work we study the time evolution 
after a quantum quench of the moments of the partial transpose, and several related quantities, 
such as the R\'enyi negativities.  
By combining Conformal Field Theory (CFT) results with integrability, we show that, 
in the space-time scaling limit of long times and large subsystems, 
a quasiparticle description allows for a complete understanding of the R\'enyi negativities. 
We test our analytical predictions against exact numerical results for 
free-fermion and free-boson lattice models, even though our framework 
applies to generic interacting integrable systems.
\end{abstract}

\newpage

\tableofcontents

\section{Introduction}

\begin{figure}[t]
\centering
\includegraphics[width=0.55\textwidth]{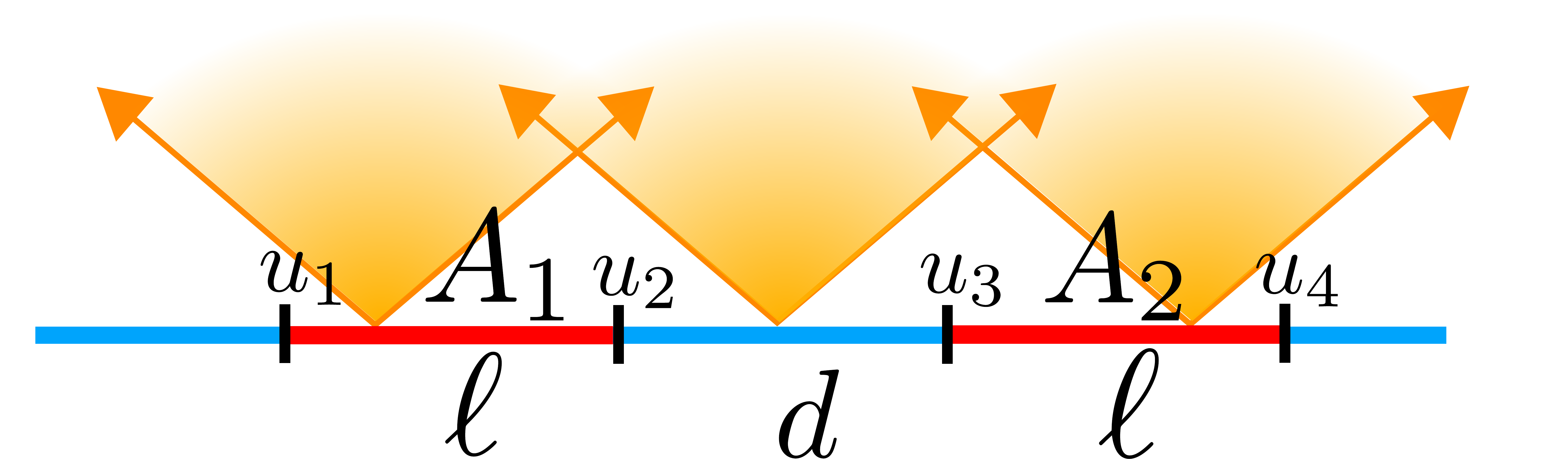}
\caption{Quasiparticle picture for the time evolution after a quench of the entanglement  between two disjoint intervals ($A_1=[u_1,u_2]$ and $A_2=[u_3,u_4]$) 
embedded in the infinite line. 
 Pairs of entangled quasiparticles are emitted from every point in space at $t = 0$.  
 At a given time $t$ the entanglement between $A$ and the remainder is proportional to the number of pairs shared between $A$ and  its complement. 
 Similarly, the entanglement between $A_1$ and $A_2$ is proportional to the pairs that are shared between them, and not between $A_1$ (or $A_2$) with 
 the rest separately.
}
\label{fig:qp-0}
\end{figure}

During the last decades, the study of entanglement became a powerful 
tool to explore the out of equilibrium dynamics of quantum systems. The 
simplest and most broadly studied protocol is the 
quantum quench \cite{cc-05,cc-07}: a system is prepared in the ground state of a translationally 
invariant Hamiltonian, and at a given time a sudden change modifies the Hamiltonian. 
In integrable systems, the entanglement dynamics after a quench is captured by a well-known 
quasiparticle picture \cite{cc-06,ac-17,ac-18b,c-20}. 
The key tenet of the quasiparticle picture is that the entanglement dynamics is described by 
the ballistic propagation of pairs of entangled excitations, which are produced after the 
quench (see Fig.~\ref{fig:qp-0}). 
An intense theoretical activity was accompanied by a remarkable experimental progress, e.g., 
to measure the many-body entanglement of non-equilibrium states~\cite{kaufman-2016,Elben2018,brydges-2019,exp-lukin}. 
For closed bipartite systems, the von Neumann and the R\'enyi entropies of reduced density matrices can be used as 
\emph{bona fide} measures of the entanglement shared between the two complementary parts. 
On the other hand, neither the entropies, nor the associated mutual information can be 
used to quantify the entanglement between two noncomplementary subsystems (see Fig.~\ref{fig:qp-0} for 
the situation with two disjoint sets in a one-dimensional system). 
The reason is that the state of the two subsystems is in general a mixed one. 
In this situation, the entanglement can be understood via the {\it partial transpose} of the reduced density matrix (RDM) which is defined as follows. 
Given the RDM $\rho_A$ of a subsystem $A=A_1 \cup A_2$ (see Fig.~\ref{fig:qp-0}), 
obtained after tracing out the rest of the system $B$ as 
$\rho_A\equiv {\rm Tr}_B\rho$, the partial transpose $\rho_A^{T_1}$ is 
obtained by taking the matrix transposition with respect to the degrees of 
freedom of one of the two subsystems (say $A_1$). 
The key point now is that the presence of negative eigenvalues in the spectrum of $\rho_A^{T_1}$ is a sufficient condition for $A_1$ and $A_2$ to be entangled \cite{peres-1996,s-00}. 
These negative eigenvalues are witnessed by the (logarithmic) negativity ${\cal E}\equiv \ln {\rm Tr} |\rho_A^{T_1}|$ \cite{vidal} 
which turns out also to be an entanglement monotone  \cite{plenio-2005}.

Unfortunately, computing the negativity or measuring it experimentally in quantum many-body 
systems is a daunting task. This fact sparked a lot of activity aiming at finding 
alternative entanglement witnesses for mixed states always starting from the partially transposed RDM. 
To this aim, several protocols to measure the moments $\mathrm{Tr}(\rho_A^{T_1})^n$ of the partial transpose have been proposed \cite{gbbb-18,csg-19,ekh-20,ncv-21} 
culminating with the actual experimental measure in an ion-trap setting using shadow tomography \cite{ekh-20,ncv-21}.
However, these moments are not direct indicators of the sign of the eigenvalues of  $\rho_A^{T_1}$ and hence of entanglement.  
Nevertheless, some linear combinations of them are sufficient conditions (known as $p_n$-PPT conditions, see below) for the presence of negative eigenvalues in the 
spectrum \cite{ekh-20,ncv-21} and so are witnesses of entanglement in mixed states. 
However, in contrast with the logarithmic negativity, for which a quasiparticle 
picture was derived in Ref.~\cite{ac-18},  results  for 
the dynamics of the moments of the partial transpose are available only for Conformal Field Theories~\cite{actc-14,wcr-15,ksr-20} (CFTs).

Here, by combining CFT and integrability, we derive the quasiparticle picture describing 
the dynamics of the moments of the partial transpose, and several related quantities, 
after a quantum quench in integrable systems. Specifically, we consider the R\'enyi 
negativities $\mathcal{E}_n$ defined as 
\begin{equation}
	\mathcal{E}_n=\ln(\mathrm{Tr}(\rho^{T_1}_A)^n). 
\end{equation}
Note that $\mathcal{E}_n$ are not proper entanglement measures, although 
the limit $\displaystyle\lim_{n_e\to1}\mathcal{E}_{n_e}$, with $n_e$ an even integer, 
defines the logarithmic negativity. We also consider the ratios $R_n$ as 
\begin{equation}
	R_n=\frac{\mathrm{Tr}(\rho_A^{T_1})^n}{\mathrm{Tr}\rho_A^n}.
\end{equation}
The ratios $R_n$ are studied in CFT~\cite{cct-12,cct-13,ctt-13,a-13,cabcl-14}, due 
to their universality. Recently, they were studied at finite-temperature 
critical points~\cite{wlckg-20}, and to probe thermalization~\cite{wkp-20,lg-20} (note that the authors of 
Ref.~\cite{lg-20} refer to the ratios $R_n$ as R\'enyi negativities, unlike here). 
Here we derive the quasiparticle picture for both $\mathcal{E}_n$ and $R_n$, focusing on the 
situation in which the subsystem $A$ is made of two equal-length intervals at distance $d$. The formulas 
that we derive hold in the space-time scaling limit of $t,\ell,d\to\infty$, with the ratios $t/\ell,d/\ell$ 
fixed. 
Furthermore, these results allow us to obtain predictions for all the $p_n$-PPT conditions 
introduced in Refs.~\cite{ekh-20,ncv-21}. Interestingly, we argue that the ratios $R_n$ in 
the space time scaling limit become proportional to the R\'enyi mutual information. 
Finally, we provide numerical benchmarks of our results for both free-fermion and 
free-boson models, although they are expected to hold for generic integrable systems. 

The paper is organised as follows. In Sec. \ref{sec:entmeasure} we review the 
definitions of some entanglement measures, i.e. R\'enyi entropy, mutual information, 
R\'enyi negativity. In particular, in section~\ref{sec:neg} we introduce the moments of the 
partial transpose and the negativities. In section~\ref{sec:mixed} we introduce the 
$p_n$-PPT conditions. In section~\ref{sec:review} we review the CFT predictions for the 
out-of-equilibrium behavior of the R\'enyi negativities. Specifically, in section~\ref{sec:twist} we 
review the representation of the R\'enyi negativities in terms of twist fields. In section~\ref{sec:cft-out} 
we derive the out-of-equilibrium behavior of the R\'enyi negativities and the ratios $R_n$ in CFTs. 
In Sec. \ref{sec:qp} we introduce the quasiparticle picture (in section~\ref{sec:qp-intro}) for the spreading of 
entanglement and negativity, generalizing it to the moments of the partial transpose in section~\ref{sec:qpn}.  
In section~\ref{sec:check} we present numerical benchmarks for free bosonic (in section~\ref{sec:hc}) 
and fermionic theories (in section~\ref{sec:ising}). In section~\ref{sec:stilt} we discuss the quasiparticle 
predictions for the $p_n$-PPT conditions. 
Finally, in section~\ref{sec:conclusions} we draw our conclusions and we discuss some possible extensions of our 
work.

\section{Entanglement measures for mixed states}
\label{sec:entmeasure}
The R\'enyi entanglement entropies are the most successful way to characterize the bipartite entanglement of a subsystem $A$  
of a many-body quantum system prepared in a pure state (see e.g. the reviews \cite{intro1,intro2,eisert-2010,intro3}), also from the experimental perspective \cite{Islam2015,kaufman-2016,Elben2018,brydges-2019,exp-lukin,vek-21}. 
Given the reduced density matrix (RDM) $\rho_A$ of a subsystem $A$, 
the R\'enyi entropies are defined as
\begin{equation}\label{eq:renyi}
    S_A^{(n)}=\frac{1}{1-n}\mathrm{ln}  \mathrm{Tr}\rho_A^n.
\end{equation}
From these,  the von Neumann entropy is obtained as the limit $n \to 1$ of Eq.~\eqref{eq:renyi} and also the entire spectrum of $\rho_A$ can 
be reconstructed \cite{cl-08}. The R\'enyi entropies in Eq.~\eqref{eq:renyi} can be very conveniently computed in field theory because 
for integer $n$, in the path-integral formalism, $\mathrm{Tr} \rho_A^{n}$ is the partition function on an $n$-sheeted Riemann surface 
$\mathcal{R}_n$ obtained by joining cyclically the $n$ sheets along the region $A$ \cite{cc-04,cc-09,cw-94}. 

For a mixed state, the entanglement entropies are no longer good measures of entanglement because they mix classical and quantum correlations (e.g. in a high temperature state, $S_A^{(n)}$ gives the extensive result for the thermal entropy that has nothing to do with entanglement). In this respect, a useful quantity to consider is the R\'enyi mutual information
\begin{equation}\label{eq:renyimutualinfo}
I^{(n)}_{A_1:A_2}\equiv S^{(n)}_{A_1}+S^{(n)}_{A_2}-S^{(n)}_{A_1\cup A_2}, 
\end{equation}
which is not a measure of the entanglement between $A_1$ and $A_2$, but for $n\to 1$ it quantifies the amount of global correlations between the two subsystems
(we mention that for $n\neq 1$, $I^{(n)}_{A_1:A_2}$ can be also negative \cite{kz-17} and a more complicated definition of mutual information must be employed \cite{sasc-21}). 

 As anticipated, we are interested here in the entanglement between two different regions, and the goal of the following section is to define the tools to compute it.

\subsection{Entanglement in mixed states and logarithmic negativity}
\label{sec:neg}
A very useful starting point to quantify mixed state entanglement is the Peres criterion \cite{peres-1996,s-00}, also known as PPT condition.
It states that given a system described by the density matrix $\rho_A$, a sufficient condition for the presence of entanglement between  
two subsystems $A_1$ and $A_2$ (with $A=A_1\cup A_2$) is that the partial transpose $\rho_A^{T_1}$ with respect to the degrees of freedom in $A_1$ 
(or equivalently $A_2$) has at least one negative eigenvalue. Let us introduce the partial transpose operation as follows. We can write down the density matrix as
\begin{equation}
\rho_A=\sum_{ijkl}\braket{e^1_i,e^2_j|\rho_A|e^1_k, e^2_l}\ket{e_i^1,e_j^2}\bra{e^1_k,e^2_l},
\end{equation}
where $\ket{e_j^1}$ and $\ket{e_k^2}$ are orthonormal bases in the Hilbert spaces $\mathcal{H}_1$ and $\mathcal{H}_2$ corresponding to the $A_1$ and $A_2$ regions, respectively. The partial transpose of a density matrix for the subsystem $A_1$ is defined by exchanging the matrix elements in the subsystem $A_1$, i.e.
\begin{equation}
\label{eq:bosonic}
\rho^{T_1}_A=\sum_{ijkl}\braket{e^1_k,e^2_j|\rho_A|e^1_i, e^2_l}\ket{e_i^1,e_j^2}\bra{e^1_k,e^2_l},
\end{equation}
In terms of its eigenvalues $\lambda_i$, the trace norm of $\rho_A^{T_1}$ can be written as
\begin{equation}
    \mathrm{Tr}|\rho_A^{T_1}|=\sum_i |\lambda_i|=\sum_{\lambda_i >0}|\lambda_i|+\sum_{\lambda_i <0}|\lambda_i|=1+2\sum_{\lambda_i<0}|\lambda_i|,
\end{equation}
where in the last equality we used the normalisation $\sum_i \lambda_i=1$. Here $\mathrm{Tr}|O|\equiv \mathrm{Tr} \sqrt{O^{\dagger}O}$ denotes the trace norm of the operator $O$.   This expression
makes evident that the negativity measures ``how much'' the eigenvalues of the partial
transpose of the density matrix are negative, a property which is the reason for the
name negativity. 
Therefore, starting from the Peres criterion, a measure of the bipartite entanglement for a general mixed state can be naturally defined as \cite{vidal}
\begin{equation}\label{eq:bos}
\mathcal{E}\equiv \mathrm{ln} \mathrm{Tr}|\rho_A^{T_1}|,
\end{equation}
which is known as {\it logarithmic negativity}. 
By considering the moments of the partial transpose RDM, one can define the 
R\'enyi negativities $\mathcal{E}_n$ as 
\begin{equation}\label{eq:renneg}
\mathcal{E}_n\equiv \mathrm{ln} \mathrm{Tr}(\rho_A^{T_1})^n. 
\end{equation}
The logarithmic negativity $\mathcal{E}$ is given by the following replica limit~\cite{cct-12,cct-13}
 \begin{equation}
 \mathcal{E}=\lim_{n_e \to 1}\mathcal{E}_{n_e},
 \end{equation}
 where $n_e$ denotes an even number $n_e=2m$ with $m$ integer.
For future convenience, we also introduce the ratios
\begin{equation}\label{eq:ratioRn}
R_n\equiv \frac{\mathrm{Tr}(\rho_A^{T_1})^n}{\mathrm{Tr}\rho_A^n}.
\end{equation}
The R\'enyi negativities $\mathcal{E}_n$ with integer $n\geq 2$ can also be measured in experiments \cite{csg-19,ekh-20,gbbb-18}, but they are not entanglement monotones.
The entanglement negativity and R\'enyi negativities have been used to characterize mixed states in various quantum systems such as in harmonic oscillator chains \cite{aepw-02,fcga-08,cfga-08,aa-08,a-08,mrpr-09,ez-14,sdhs-16,dnct-16}, 
quantum spin models \cite{wvb-09,bbs-10,bsb-10,bbsj-12,wvb-10,skb-11,lg-19,rac-161,trc-19,glen,waa-20,cabcl-14,a-13,slkv-21}, 
(1+1)d conformal and integrable field theories 
\cite{cct-12,cct-13,ctt-13,cabcl-14,a-13,ctc-14,rac-16,act-18,kpp-14,dct-15,bca-16,bfcad-16,cdds-19,cdds-19b,kr-19,asv-21,csg-19,mbc-21}, 
out-of-equilibrium settings \cite{ekh-20,actc-14,vez-14,hd-15,ac-18,wcr-15,gh-19,kkr-21,sdl-21,wkp-20,kkr-20,lg-20}.

Crucially, while for free-boson models the R\'enyi negativities for arbitrary real $n$ can be efficiently computed from the two-point correlation function~\cite{aepw-02}, 
this is not the case for free-fermion systems. 
The main problem is that the partial transpose in Eq.~\eqref{eq:bos} is not a gaussian operator, although it
can be written as the sum of two gaussian (non-commuting) operators $O_{\pm}$ as~\cite{ez-17}
\begin{equation}\label{eq:rhoT2}
\rho^{T_1}_A=\frac{1-i}{2}O_++\frac{1+i}{2}O_-.
\end{equation}
From this observation a procedure to extract the R\'enyi negativities of integer order was proposed  \cite{ez-17} and was also used in many subsequent studies \cite{ctc-15,ctc-16,acetpc-16,ez2d,cxw-16,hw-16,eez-18}. 
Still, proceeding in this way, it is not possible to perform the replica limit $n_e\to1$, implying that the negativity, i.e. the only genuine measure of entanglement, is not accessible.
To overcome this problem, an alternative estimator of mixed-state entanglement for fermionic systems has been 
introduced based on the time-reversal partial transpose (a.k.a  {\it partial time reversal}) \cite{ssr-17,ssr1-17,ssgr-18,ryu,paola,smr-21,sr-19,ksr-20}.
The new estimator has been dubbed {\it fermionic negativity}. 
It has been shown that the fermionic negativity is an entanglement monotone \cite{sr-19} and it is also an upper bound for the standard
negativity. In the following, we denote $\mathcal{E}_n^{(b)}$
the standard negativity in Eq.~\eqref{eq:renneg} (that is an entanglement monotone for both bosonic and fermionic systems) and $\mathcal{E}_n^{(f)}$ the fermionic 
one (that exist only for fermionic models).
$\mathcal{E}^{(f)}$ reads as 
\begin{equation}
\mathcal{E}^{(f)}= \mathrm{ln} \mathrm{Tr} \sqrt{O_+O_-},
\label{NF}
\end{equation}
with $O_\pm$ as defined implicitly in Eq.~\eqref{eq:rhoT2}. 
The (fermionic) R\'enyi negativities can be defined as \cite{ssr-17}
\begin{equation}\label{eq:sup}
\mathcal{E}_n^{(f)}= 
\begin{cases}
  \mathrm{ln}[\mathrm{Tr}(O_+O_-\dots  O_+O_-], &\quad n \quad \mathrm{even},\\
 \mathrm{ln}[\mathrm{Tr}(O_+O_-\dots O_+)],& \quad n \quad \mathrm{odd},
\end{cases}
\end{equation}
from which $  \displaystyle \mathcal{E}^{(f)}=\lim_{n_e \to 1}  \mathcal{E}^{(f)}_{n_e}$.
The products involving $O_+$ and $O_-$ are still gaussian fermionic operators, so all the above quantities can be efficiently computed, 
including the negativity \eqref{NF}.

 \subsection{Entanglement detection through partial transpose moments}
 \label{sec:mixed}
 Despite several sufficient conditions for entanglement in mixed states have been developed in the literature,
many of them cannot be straightforwardly implemented experimentally since they require the knowledge of the full density matrix \cite{intro1}.
This is for instance the case of the PPT condition.  To overcome this difficulty, it
was shown in \cite{ekh-20} that the first few moments of the partial transpose can be used to define some simple yet powerful tests
for bipartite entanglement. Given $\rho_A^{T_1}$ (cf. Eq.~\eqref{eq:bosonic}), we denote its $k$-th order moment as
\begin{equation}\label{eq:pmoments}
p_k\equiv \mathrm{Tr}(\rho_A^{T_1})^k,
\end{equation}
with $p_1 = \mathrm{Tr}(\rho^{T_1}_A) = 1$ and $p_2$ equal to the purity $p_2 = \mathrm{Tr}\rho_A^2$.
The $p_3$-PPT condition states that any positive semi-definite partial transpose satisfies~\cite{ekh-20}
\begin{equation}
	\label{eq:p3}
p_3 p_1 > p_2^2,
\end{equation}
or, in other words, if $p_3 < p_2^2$, then $\rho_A$ violates the PPT condition and must therefore be entangled. 
The condition in Eq.~\eqref{eq:p3} belongs to a more general set of conditions, dubbed Stieltjes$_n$, involving inequalities among the moments $p_k$ of order up to $n$. 
They were introduced in \cite{ncv-21} together with a set of experimentally accessible conditions for detecting entanglement in mixed states.
The condition Stieltjes$_3$ is equivalent to $p_3-$PPT, and so we rename here the Stieltjes$_n$-conditions as $p_n$-PPT.
As examples, $p_5$-PPT and $p_7$-PTT read, respectively \cite{ncv-21}
\begin{equation}
D_5\equiv\det \begin{pmatrix} 
    p_{1}&p_{2}&p_{3} \\
    p_{2}&p_{3}&p_{4} \\
    p_{3}&p_{4}&p_{5} 
    \end{pmatrix}\geq 0,\quad
D_7\equiv    \det \begin{pmatrix} 
    p_{1}&p_{2}&p_{3}&p_{4} \\
    p_{2}&p_{3}&p_{4}&p_{5} \\
    p_{3}&p_{4}&p_{5} &p_{6} \\
    p_{4}&p_{5}&p_{6} &p_{7} 
    \end{pmatrix}\geq 0,
    \label{DDn}
\end{equation}
from which one deduces easily the rationale for higher order condition. 

\section{Quench dynamics of R\'enyi negativities in conformal field theory}
\label{sec:review}
In this section we review the CFT calculation of the temporal evolution of the R\'enyi negativities 
between two intervals after a global quench in CFT as derived in Ref. \cite{actc-14}. We consider 
$A = A_1 \cup A_2$, where the intervals $A_1$ and $A_2$ can be either adjacent or disjoint (see Fig.~\ref{fig:qp-0}).

\subsection{R\'enyi negativities from twist field correlation functions}
\label{sec:twist}
A powerful method to calculate the R\'enyi negativities is based on a particular type of twist fields 
in quantum field theory that are related to branch points in the Riemann surface 
$\mathcal{R}_n$ \cite{cc-09,ccd-08}. We denote twist and anti-twist fields by 
$\mathcal{T}_n$ and $\mathcal{\tilde{T}}_n$, respectively. One can show that the moments of the 
reduced density matrix $\mathrm{Tr}(\rho_A)^n$ can be written as correlators of twist fields~\cite{cc-09}. 
For example, when $A=[u_1,u_2]\cup [u_3,u_4]$ (see Fig.~\ref{fig:qp-0}), one has that  
\begin{equation}
	\label{eq:exp}
\mathrm{Tr}\rho_A^n = \braket{\mathcal{T}_n(u_1)\mathcal{\tilde{T}}_n(u_2)\mathcal{T}_n(u_2)\mathcal{\tilde{T}}_n(u_3)}.
\end{equation} 
Notice that the twist and anti-twist fields are inserted at the endpoints of $A$.  
The expectation value in Eq.~\eqref{eq:exp} is taken with respect to the action living on a plane. 
As shown in \cite{cct-13}, if we take the partial transpose $\rho_A^{T_1}$ with respect to the degrees of freedom 
living on the interval $A_1=[u_1,v_1]$, $\mathrm{Tr}(\rho^{T_1}_A)^n$ can be written as a twist-field correlator as in Eq.~\eqref{eq:exp}, 
the only difference being that the twist fields $\mathcal{T}_n$ and $\mathcal{\tilde{T}}_n$ at the endpoints of 
$A_1$ are exchanged while the remaining ones stay untouched, i.e. \cite{cct-12,cct-13}
\begin{equation}
	\label{eq:exp-1}
\mathrm{Tr}(\rho^{T_1}_A)^n = \braket{\mathcal{\tilde{T}}_n(u_1)\mathcal{T}_n(u_2)\mathcal{T}_n(u_3)\mathcal{\tilde{T}}_n(u_4)}.
\end{equation} 
This procedure can be generalized straightforwardly to the case where $A$ is the union of more than two intervals, 
and the partial transposition  involves more than two intervals. 

The situation in which the two intervals are adjacent can be obtained from Eq.
\eqref{eq:exp-1} by taking the limit $u_3\to u_2$ in Eq.~\eqref{eq:exp-1}, giving  
$\mathrm{Tr}(\rho^{T_1}_A)^n = \braket{\mathcal{\tilde{T}}_n(u_1)\mathcal{T}^2_n(u_2)\mathcal{\tilde{T}}_n(u_4)}
$. 
In a generic CFT characterized by a central charge $c$, the expectation 
values~\eqref{eq:exp} and~\eqref{eq:exp-1} are evaluated straightforwardly by knowing 
the scaling dimensions of $\mathcal{T}_n,\mathcal{\tilde{T}}_n$, $\mathcal{T}^2_n$ 
and $\mathcal{\tilde{T}}^2_n$.  The scaling dimensions of $\mathcal{T}^2_n$ and 
$\mathcal{\tilde{T}}^2_n$ are equal, and  depend on the parity of $n$ as~\cite{cct-12} 
\begin{equation}
	\label{eq:D2}
\Delta^{(2)}_n\equiv\begin{cases}
\Delta_n \qquad \mathrm{odd}\, n \\
2\Delta_{n/2} \quad \mathrm{even}\, n \\
\end{cases}, \quad \Delta_n =\frac{c}{12}\left(n-\frac{1}{n} \right), 
\end{equation}
where $\Delta_n$ are the scaling dimensions of $\mathcal{T}_n,\mathcal{\tilde{T}}_n$. 

\subsection{Out-of-equilibrium dynamics of the R\'enyi negativities }
\label{sec:cft-out}

Before discussing the out-of-equilibrium dynamics after a quantum quench of the R\'enyi negativities 
in CFTs, it is useful to recall  the imaginary time formalism for the description of quantum quenches~\cite{cc-05,cc-06,cc-07}. 
The family of initial states that are easy to work with in CFT have the form $e^{-\tau_0 H} |\psi_0\rangle$, with $ |\psi_0\rangle$ being a boundary state.
The expectation value of a local operator $O$ is then 
\begin{equation}\label{eq:corr}
\braket{O(t,x)} = Z^{-1}\bra{\psi_0}e^{iHt-\tau_0H}O(x)e^{-iHt-\tau_0H}\ket{\psi_0},
\end{equation}
where the damping factors $e^{-\tau_0H}$ have been introduced to make the path integral absolutely
convergent (see below), and $Z=\braket{\psi_0|e^{-2\tau_0H}|\psi_0}$ is the normalisation 
factor. The correlator in Eq.~\eqref{eq:corr} may be represented by a path integral in imaginary time $\tau$ as~\cite{cc-07}
\begin{equation}
	\label{eq:pi}
\braket{O(t,x)} = Z^{-1}\int [d\varphi(x,\tau)]O(x,\tau=\tau_0+it)e^{-\int_{\tau_1}^{\tau_2}Ld\tau}\braket{\psi_0|\phi(x,\tau_2)\rangle \langle \phi(x,\tau_1)|\psi_0},
\end{equation}
where $L$ is the (euclidean) Lagrangian corresponding to the dynamics induced by 
$H$, $\tau_1$ can be identified with $ 0$ and $\tau_2$ with  $ 2\tau_0 $. As shown 
in \cite{cc-05,cc-06}, the computation of the path integral in Eq.~\eqref{eq:pi} can 
be done considering $\tau$ real and only at the end analytically continuing it to 
the complex value $\tau = \tau_0 + it$. 

\begin{figure}[t]
\centering
\includegraphics[width=0.95\textwidth]{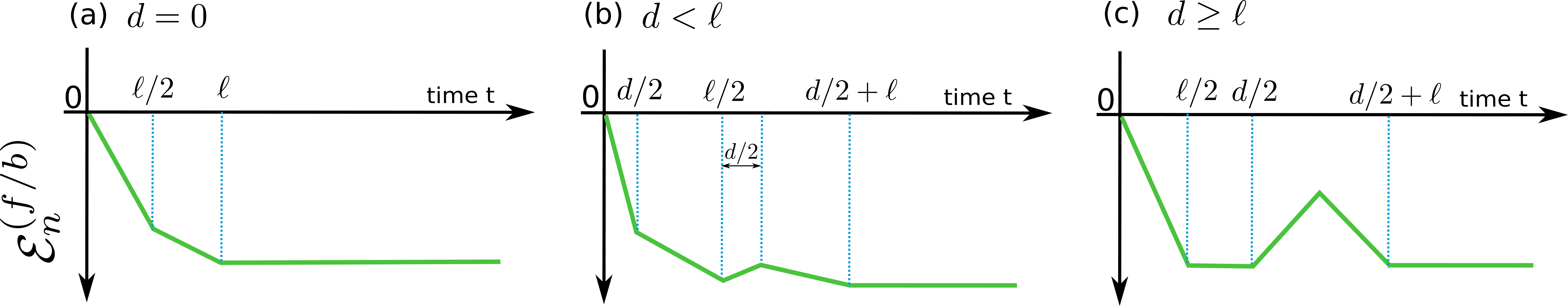}
\caption{
Time dependence of the R\'enyi negativities for quasiparticles with linear dispersion. 
We consider two disjoint subsystems with equal length $\ell$ at distance $d=0$ (a), $d<\ell$ (b), $d\geq \ell$ (c). 
The piece-wise linear behavior is described by Eq.~\eqref{eq:quasi1}. 
}
\label{fig:qp}
\end{figure}

In this way, the problem of the dynamics is mapped to the thermodynamics 
of a field theory in a strip geometry of width $2\tau_0$ and boundary condition 
$|\psi_0\rangle$ at the two edges of the strip in the imaginary time direction. 
At this point we have all the ingredients to derive  the dynamics  of
the R\'enyi negativities after a global quench in CFT. 
To calculate the time-dependent $\mathrm{Tr}(\rho_A^{T_1})^n$ one has to compute the correlator
\begin{equation}
	\label{eq:tr-twist}
\mathrm{Tr}(\rho^{T_1}_A)^n = \braket{\mathcal{\tilde{T}}_n(\omega_1)\mathcal{T}_n(\omega_2)\mathcal{T}_n(\omega_3)\mathcal{\tilde{T}}_n(\omega_4)}, 
\end{equation} 
where the expectation value has to be calculated in the field theory confined in 
a strip, and where we denoted by $\omega_i = u_i+i\tau$ the complex coordinate on 
the strip ($u_i \in \mathbb{ R}$ and $0<\tau<2\tau_0$). 
It is convenient to employ the  conformal transformation $z=e^{\pi \omega/(2\tau_0)}$, which 
maps the strip onto the half plane, where the four point correlation functions of the twist fields can be 
computed by knowing that they behave as primary fields with scaling dimenions $\Delta_n$ (cf.~\eqref{eq:D2}).  
After the analytic continuation to real  time, in the space-time scaling limit $t,|u_i-u_j|\gg \tau_0$, 
from Eq.~\eqref{eq:tr-twist}, the R\'enyi negativities $\mathcal{E}_n$ (cf. Eq.~\eqref{eq:renneg}) read~\cite{actc-14}
\begin{multline}\label{eq:quasi1}
\mathcal{E}_n=-\frac{\pi}{\tau_0}\Big[2\Delta_nt+\Delta_n\left(\frac{\ell_1+\ell_2}{2}-\mathrm{max}(t,\ell_1/2)-\mathrm{max}(t,\ell_2/2)\right)+(\Delta^{(2)}_n/2-\Delta_n)\\ \times\left(\mathrm{max}(t,(\ell_1+\ell_2+d)/2)+\mathrm{max}(t,d/2)\right)-\mathrm{max}(t,(\ell_1+d)/2)-\mathrm{max}(t,(\ell_2+d)/2)\Big],
\end{multline}
where $\Delta_n^{(2)}$ is in Eq.~\eqref{eq:D2}, and we defined $\ell_1=|u_1-u_2|$, $\ell_2=|u_3-u_4|$, and $d=|u_3-u_2|$ 
(see Fig.~\ref{fig:qp-0}). 
In deriving Eq.~\eqref{eq:quasi1} we neglected an additive time-independent constant that 
originates from the correlation function of the twist fields, and that depends on the details of the CFT under consideration. 
This is justified because Eq.~\eqref{eq:quasi1} holds in the scaling limit $\ell_1,\ell_2,d,t\to\infty$ with their 
ratios fixed, and it describes only the leading behavior of the R\'enyi negativities $\mathcal{E}_n$ in that limit. 
Crucially, Eq.~\eqref{eq:quasi1} depends on both $\Delta_n$ and $\Delta_n^{(2)}$. Notice that 
even for finite $d$, $\mathcal{E}_n$ exhibits a linear behavior at short times, due to the first term in Eq.~\eqref{eq:quasi1}. This 
signals that $\mathcal{E}_n$ are not good measures of the entanglement or the correlation between $A_1$ and $A_2$. 
The reason is that for $t\ll d$ no correlation can be shared between $A_1$ and $A_2$ because the maximum velocity in the 
system is finite (see Fig.~\ref{fig:qp}). 
We stress that Eq.~\eqref{eq:quasi1} is not directly applicable to microscopic integrable models: Eq.~\eqref{eq:quasi1} 
is only valid for CFT, in which there is a perfect linear dispersion, i.e., only one velocity. 
This is not the case in integrable lattice models, where the excitations have a nonlinear dispersion. 
In the next sections, we will show how to adapt Eq.~\eqref{eq:quasi1} to describe the dynamics of the R\'enyi negativities 
after a quantum quench in microscopic integrable systems. 

Finally, the dynamics of the ratio $R_n$ in Eq.~\eqref{eq:ratioRn}  can be derived combining Eq. \eqref{eq:quasi1} 
with the results for ${\rm Tr}\rho_A^n$ in \cite{cc-06}. The final result reads~\cite{actc-14} 
\begin{multline}\label{eq:quasi2}
\mathrm{ln} R_n=\frac{\pi \Delta_n^{(2)}}{\tau_0} \\ \times \left(-\mathrm{max}(t,(\ell_1+\ell_2+d)/2)-\mathrm{max}(t,d/2)\right)
+\mathrm{max}(t,(\ell_1+d)/2)+\mathrm{max}(t,(\ell_2+d)/2).
\end{multline}
In contrast with Eq.~\eqref{eq:quasi1}, Eq.~\eqref{eq:quasi2}  
does not depend explicitly on $\Delta_n$, but only on $\Delta_n^{(2)}$.

\begin{figure}[t]
\centering
\includegraphics[width=0.95\textwidth]{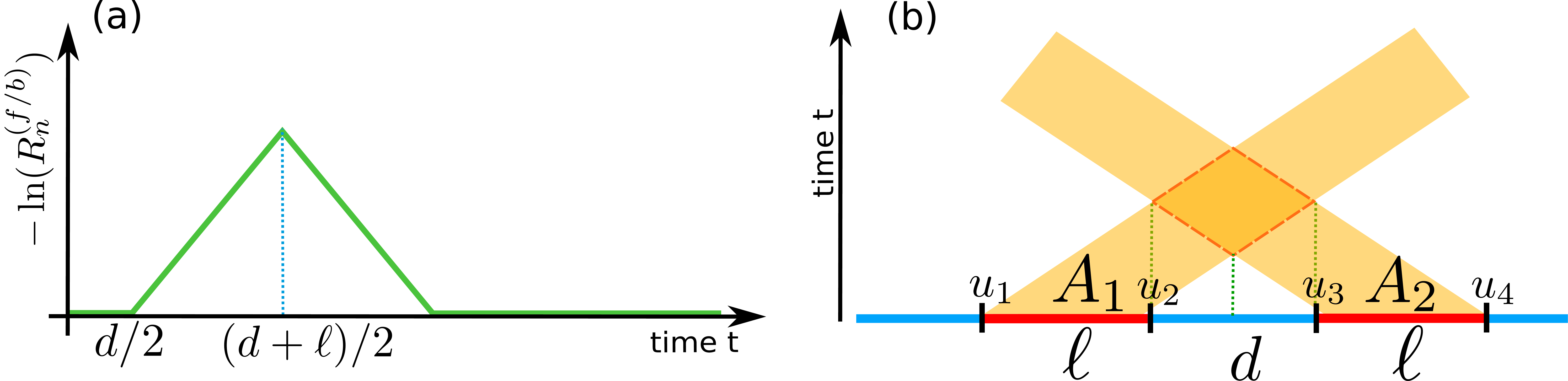}
\caption{
 Illustration of the dynamics of $-\mathrm{ln} R_n^{(f/b)}$ for two disjoint subsystems with equal length $\ell$ at distance $d$.
On the left we report the shape of $-\mathrm{ln} R_n^{(f/b)}$ with a single velocity of quasiparticles. On the right, there is a graphical representation for the quasi-particle spreading of entanglement (for the case with all quasi-particles having the same velocity $v = 1$ as in a CFT). 
Horizontal slices of the dark orange region count the quasiparticles shared between the two disjoint sets at a given time. 
}
\label{fig:qp-1}
\end{figure}

Before concluding, it is useful to discuss the qualitative behavior of $\mathcal{E}^{(f/b)}_n$ and 
$-\ln(R_n^{(f/b)})$.  The typical behavior of the R\'enyi negativities, as obtained from Eq.~\eqref{eq:quasi1},  is reported 
in Fig.~\ref{fig:qp} for three typical values of the distance $d$ between the two intervals of equal length $\ell$. 
$\mathcal{E}_n$ is always a piecewise linear function and it is negative at any time. 
For $d=0$ one has a two-slope linear behavior followed by a saturation to a volume-law scaling at long times. 
At intermediate distance $0<d<\ell$ the behavior is more 
complicated with a change in the sign of the slope. For $d>\ell$, $\mathcal{E}_n^{(f/b)}$ 
exhibits an initial linear decrease followed by a saturation, and a dip-like feature at $d/2\le t\le d/2+\ell$. 

The dynamics of $-\mathrm{ln} (R_n^{(f/b)})$ (cf. Eq.~\eqref{eq:quasi2}) is shown  in Fig.~\ref{fig:qp-1}  for two equal-length intervals. 
For $t < d/2$,
it vanishes; for $d/2 \leq t \leq (d + \ell)/2$ it linearly increases, then it linearly decreases with the same
(in absolute value) slope until $t \leq (d + 2\ell)/2$, when it vanishes and stays zero for all larger times. 
Therefore, at a given time $t$, it is proportional to the
width of the intersection between the two shaded areas starting from $A_1 \cup A_2$ 
and showed in Fig.~\ref{fig:qp-1} (b). In other words,  it is proportional to 
the total number of entangled pairs shared between $A_1$ and $A_2$. 
This property suggests that in the scaling limit, $R_n$ becomes an indicator of
the mutual entanglement between the intervals, although in general it is not an entanglement monotone. 
 
Let us remark that Eq.~\eqref{eq:quasi2}  is identical to the evolution of the R\'enyi mutual
information in Eq.~\eqref{eq:renyimutualinfo} apart from the prefactor. 
We will come back to the connection between these two quantities in the following sections.

\section{Quasiparticle picture for the R\'enyi negativities in integrable systems}
\label{sec:qp}

The goal of this section is to adapt Eq.~\eqref{eq:quasi1} and Eq.~\eqref{eq:quasi2} to 
describe the dynamics of the R\'enyi negativities and the ratios $R_n$ after a quantum quench in integrable systems. 
The main observation is that Eq.~\eqref{eq:quasi1} and Eq.~\eqref{eq:quasi2} admit an interpretation in terms of a
simple hydrodynamic picture, a.k.a. the quasiparticle picture. 

\subsection{Quasiparticle picture}
\label{sec:qp-intro}

The quasiparticle picture for the entanglement dynamics after a global quantum quench has been 
proposed in Ref.~\cite{cc-06}. The underlying idea is that the pre-quench 
initial state has very high energy with respect to the ground state of the Hamiltonian 
governing the dynamics; hence it can be seen as a source of quasiparticle excitations at $t = 0$.  
We assume that quasiparticles are uniformly created in uncorrelated pairs with quasimomenta $(k,-k)$ and 
traveling with opposite velocities $v(k)=-v(-k)$ (for free models the uncorrelated pair assumption can be released, see \cite{btc-17,btc-18,bc-18,bc-20b}; 
for interacting integrable models it has been argued that the pair structure is what makes the quench integrable \cite{PPV-17}). 
Quasiparticles produced at the same point in space are entangled,
whereas quasiparticles created far apart are incoherent. The quasiparticles travel through the 
system as free-like excitations. At a
generic time $t$, the von Neumann entropy and the R\'enyi entropies 
between a subsystem $A$ and the rest is proportional to the total number of
quasiparticles that were created at the same point at $t = 0$ and are 
shared between $A$ and its complement at time $t$ (see Fig. \ref{fig:qp} (a)). 
Let us focus on the quasiparticle picture for the R\'enyi entropies in free models  
(the quasiparticle picture has been derived rigorously for free-fermion models in Ref.~\cite{fagotti-2008}). 
In formulas it reads as 
\begin{equation}
	\label{eq:sn-ff}
	S^{(n)}_A(t)=\int\frac{dk}{2\pi}s_{GGE}^{(n)}(k)\min(2|v(k)|t,\ell). 
\end{equation}
Here  $\ell$ is the length of subsystem $A$, and $v(k)$ is the group velocity 
of the fermionic excitations. Importantly, in Eq.~\eqref{eq:sn-ff} $s_{GGE}^{(n)}(k)$ is the density (in momentum space) of the 
R\'enyi entropies of the GGE thermodynamic state~\cite{cem-16,vr-16,ef-16} that describes the steady state after the quench. 
Eq.~\eqref{eq:sn-ff} predicts a linear growth for $t\le\ell/(2v_\mathrm{max})$, with $v_\mathrm{max}\equiv \max_k (v(k))$ 
the maximum velocity in the system, and then saturates to an extensive value at $t\to\infty$. 

For $n=1$, i.e., for the von Neumann entropy the validity of Eq.~\eqref{eq:sn-ff} for a generic interacting integrable model has been conjectured in 
Ref.~\cite{ac-17,ac-18b}. Eq.~\eqref{eq:sn-ff} remains essentially the same. Precisely, the contribution of the 
quasiparticles to the von Neumann entropy $s^{(1)}_{GGE}$ is the density of GGE thermodynamic entropy. 
The group velocities of the quasiparticles are obtained as particle-hole excitations over the GGE thermodynamic macrostate~\cite{bonnes-2014,ac-18b}. 
This conjecture has been explicitly worked out in several cases \cite{ac-17,ac-18b,pvcp-18,mbpc-17} and tested against numerics in several interacting integrable models. \cite{ac-17,ac-18b,mpc-19,pvcr-16}
Eq.~\eqref{eq:sn-ff} has been generalized to describe the steady-state value of the 
R\'enyi entropies~\cite{renyi,renyi-1,mestyan-2018}. On the other hand, the full-time dynamics of the 
R\'enyi entropies is still an open problem, with the exception of one model~\cite{klobas-2021,klobas-2021a}. 
Eq.~\eqref{eq:sn-ff} can be straightforwardly generalized to describe 
the dynamics of the mutual information between two intervals. This allows to reveal how quantum information 
is scrambled in integrable systems~\cite{ac-19,mac-20}. 
Remarkably, the quasiparticle picture for the 
logarithmic negativity has been derived in Ref.~\cite{act-18}. By combining the quasiparticle picture with 
the framework of the Generalized Hydrodynamics~\cite{cdy-16,bcnf-16} it is possible to describe the entanglement dynamics 
after quenches from inhomogeneous initial states~\cite{alba-2018a,bertini-2018,alba-2019,alba-2019a,mestyan-2020,alba-rev}. 
The quasiparticle picture for the entanglement dynamics has been also tested in the  rule $54$ chain, 
which is believed to be a representative ``toy model'' for generic interacting integrable systems~\cite{klobas-2021,klobas-2021a}. 
Very recently, the quasiparticle picture has been generalized to take into account dissipative effects, at least in free-fermion 
and free-boson models~\cite{alba-d,carollo-d,alba-da,a-21}, to describe the evolution of the  symmetry-resolved 
entanglement entropies~\cite{pbc-20,pbc-21}, and for the characterization of the prethermalization dynamics \cite{bc-20}.

To proceed it is useful to compare Eq.~\eqref{eq:sn-ff} with the CFT prediction for the 
dynamics of the R\'enyi entropies~\cite{cc-06} 
\begin{equation}
	\label{eq:renyi-d-cft}
	S_A^{(n)}=-\frac{1}{1-n}\frac{\pi\Delta_n}{2\tau_0}\min(2t,\ell). 
\end{equation}
A crucial observation is that Eq.~\eqref{eq:sn-ff} can be formally obtained from the 
CFT result in Eq.~\eqref{eq:renyi-d-cft} by replaing $t\to |v(k)|t$, integrating over the quasiparticles with quasimomentum 
$k$, and replacing $-\pi\Delta_n/(2\tau_0)\to s_{GGE}^{(n)}$.

\subsection{The quasiparticle description for R\'enyi negativities}\label{sec:qpn}
The quasiparticle picture described above can be adapted to describe the R\'enyi negativities 
$\mathcal{E}_n^{(f/b)}$ and the ratios $\ln(R_n^{(f/b)})$, 
in integrable systems after a global quench. 

Indeed, similarly to the R\'enyi entropies, from Eqs.~\eqref{eq:quasi1} and~\eqref{eq:quasi2}, by using Eq.~\eqref{eq:D2}, 
after replacing $-\pi\Delta_n/(2\tau_0)\to s_{GGE}^{(n)}$, and by integrating over $k$, one obtains that 
\begin{multline}\label{eq:enwd}
\mathcal{E}^{(f/b)}_n=\int \frac{dk}{2\pi}\Big[ 4\varepsilon_n |v|t+2\varepsilon_n\left(\frac{\ell_1+\ell_2}{2}-\mathrm{max}(|v|t,\ell_1/2)-\mathrm{max}(|v|t,\ell_2/2)\right)\\
-(2\varepsilon_n-\varepsilon_n^{(2)})\big(\mathrm{max}(|v|t,(\ell_1+\ell_2+d)/2)+\mathrm{max}(|v|t,d/2) \\-\mathrm{max}(|v|t,(\ell_1+d)/2)-\mathrm{max}(|v|t,(d+\ell_2)/2)\big)\Big],
\end{multline}
 while the ratios $R_n^{(f/b)}$ read
 \begin{multline}\label{eq:rn}
\mathrm{ln}(R^{(f/b)}_n)=\int \frac{dk}{2\pi} \varepsilon_n^{(2)}\big(\mathrm{max}(|v|t,d/2)-\mathrm{max}(|v|t,(\ell_1+d)/2) \\-\mathrm{max}(|v|t,(\ell_2+d)/2)+\mathrm{max}(|v|t,(\ell_1+\ell_2+d)/2)\big).
\end{multline}
We defined 
\begin{equation}
	\label{eq:map}
\varepsilon_n^{(2)}(k)\equiv \begin{cases}
\varepsilon_n(k) \qquad &{\rm odd} \, n\\
2\varepsilon_{n/2}(k) \qquad &{\rm even} \, n
\end{cases},\quad \varepsilon_n(k)=s_{GGE}^{(n)}(k). 
\end{equation}
Clearly, Eq.~\eqref{eq:map} mirrors the structure of Eq.~\eqref{eq:D2}. Here $s_{GGE}(k)$ is the 
density of $GGE$ thermodynamic entropy. 

It is interesting to remark that by comparing Eq.~\eqref{eq:rn} with the quasiparticle picture 
for the R\'enyi mutual informations~\cite{ac-18b} $I_{A_1:A_2}^{(n)}$, one obtains 
\begin{equation}\label{eq:pure1}
\ln(R_n^{(f/b)})=\begin{cases}
(1-n)I^{(n/2)}_{A_1:A_2} \quad n \mbox{ even}\\ \\
(1-n)\frac{I^{(n)}_{A_1:A_2}}{2},\quad n\mbox{ odd}.
\end{cases}
\end{equation}
Moreover, by taking the replica limit $n_e\to 1$ in $\mathcal{E}_{n_e}^{(f/b)}$, we recover the 
quasiparticle prediction for the negativity \cite{ac-18}
\begin{multline}\label{eq:negalba}
\mathcal{E}^{(f/b)}=\int \frac{dk}{2\pi} \varepsilon_{1/2}(k)\big(\mathrm{max}(2|v|t,d)-\mathrm{max}(2|v|t,\ell_1+d) \\-\mathrm{max}(2|v|t,\ell_2+d)+\mathrm{max}(2|v|t,\ell_1+\ell_2+d)\big).
\end{multline}
It was pointed out in \cite{ac-18} that Eq.~\eqref{eq:negalba} is the same as for the R\'enyi mutual information (of any index) by replacing $\varepsilon_{1/2}$ with the density of R\'enyi entropy. 
We stress that the same prediction is valid for both standard (bosonic) partial transpose and for the fermionic one.

Finally, it is useful to observe that Eq.~\eqref{eq:pure1} can be derived by 
using that if $A_1\cup A_2$ is in a pure state then $\mathrm{Tr}((\rho_A^{T_1})^n)$ can be 
expressed in terms of $\mathrm{Tr}(\rho_{A_1}^n)$. More precisely, one can prove that~\cite{cct-13}  
\begin{equation}\label{eq:pure}
\mathrm{Tr}(\rho_A^{T_1})^n=\begin{cases}
	\mathrm{Tr}\rho_{A_1}^{n} \,  \quad n\mbox{ odd} \\
	(\mathrm{Tr}\rho_{A_1}^{n/2})^2 \, \quad n\mbox{ even} 
\end{cases}
\end{equation}
where $\rho_{A_1}=\mathrm{Tr}_{A_2}\rho_A$. 
Now, one can recover Eq.~\eqref{eq:pure1} by using Eq.~\eqref{eq:pure}, and the definition in 
Eq.~\eqref{eq:renyimutualinfo}, and that if $A_1\cup A_2$ is in a pure state, 
$S^{(n)}_{A_1}=S^{(n)}_{A_2}$. The fact that the result of the quasiparticle picture~\eqref{eq:quasi1} is not 
sensitive to $A_1\cup A_2$ not being in a pure state reflects that the initial state has low entanglement and 
that during the dynamics the entanglement is transported ballistically. 

Finally, for Eq.~\eqref{eq:enwd} and Eq.~\eqref{eq:rn} to be predictive one has to fix the function $s_{GGE}(k)$  (cf. Eq.~\eqref{eq:map}). 
Here we focus on out-of-equilibrium protocols for free-fermion and free-boson models. In this situation, $s_{GGE}(k)$ is determined from 
the  population of the modes $\rho(k)$ of the postquench Hamiltonian in the stationary state (see Refs.~\cite{c-18,c-20} for a pedagogical review). 
Actually, since $\rho(k)$ are conserved they can be equivalently computed in the initial state, without solving the dynamics. 
Specifically, one has that 
\begin{equation}
s_{\mathrm{GGE}}^{(n,f/b)}(k)=\pm\ln(\pm\rho(k)^n+(1\mp \rho(k))^{n}),
\label{sGGE}
\end{equation}
where the upper and lower signs are for fermionic and bosonic systems, respectively (and not to fermionic and bosonic negativity). 
We remark  that, although the quasiparticle prediction in Eqs. \eqref{eq:enwd} and \eqref{eq:rn} is expected to be 
valid also for interacting integrable models, the full quasiparticle picture for the R\'enyi entropies is not known.

\section{Time evolution of R\'enyi negativities in free models: Numerical results}\label{sec:check}

In this section we provide numerical benchmarks for the results of section~\ref{sec:qpn}. As an example of 
free-bosonic system, we consider the harmonic chain. Our results for free-fermion systems are tested 
against exact numerical data for a fermionic chain. 

\subsection{Mass quench in the harmonic chain}
\label{sec:hc}

Let us start discussing the dynamics of the R\'enyi negativities after a mass quench in the 
harmonic chain. The harmonic chain is described by the Hamiltonian 
\begin{equation}
	\label{eq:hc-ham}
	H=\frac{1}{2}\sum_{n=0}^{L-1}p^2_n+m^2q^2_n+(q_{n+1}-q_n)^2,  \quad q_0 = q_L , \, p_0 = p_L,
\end{equation}
where $L$ is the number of lattice sites, $q_n$ and $p_n$ are
canonically conjugated variables, with $[q_n, p_m] = i\delta_{nm}$, 
and $m$ is a mass parameter. The harmonic chain can be diagonalized 
in Fourier space and is equivalent to a system of free bosons. 
The dispersion relation of the bosons is given by~\cite{c-20} 
\begin{equation}
e(k) = [m^2 + 2(1 - \cos (k))]^{1/2}.
\end{equation}  
The group velocities are obtained from the single particle energies $e(k)$ 
as 
\begin{equation}
v(k) = \frac{de(k)}{dk}=\sin(k)[m^2 + 2(1 - \cos (k))]^{-1/2},
\end{equation} 
and the maximum one is $v_{\mathrm{max}}=\mathrm{max}_kv(k)$. In the mass 
quench protocol, the system is prepared in the ground state $\ket{\psi_0}$ of the 
Hamiltonian~\eqref{eq:hc-ham} with $m=m_0$. At $t = 0$ the mass parameter is
quenched from $m_0$ to a different value $m$ and the system unitarily evolves under 
the new Hamiltonian $H(m)$, namely $\ket{\psi(t)}=e^{-iH t}\ket{\psi_0}$. 
The density $\rho(k)$ (cf. Eq.~\eqref{sGGE}) of the bosons 
is written in terms of the pre- and post-quench dispersions 
$e_0(k)$ and $e(k)$ as~\cite{cc-06,cc-07,c-20}
\begin{equation}
\rho(k)=\frac{1}{4}\left(\frac{e(k)}{e_0(k)}+\frac{e_0(k)}{e(k)} \right)-\frac{1}{2}.
\end{equation}
For free bosonic systems the R\'enyi negativities can be  constructed
from the two-point correlation functions $\langle q_i q_j\rangle$, $\langle p_i p_j\rangle$ and $\langle q_i p_j\rangle$. 
Indeed, given a subsystem $A$ containing $\tilde{\ell}$ sites, which could
be either all in one interval or in disjoint intervals, the reduced density matrix
for $A$ can be studied~\cite{ep-09,actc-14} by constructing the $\tilde{\ell}\times \tilde{\ell}$ 
matrices $Q^A_{ij} = \braket{q_i q_j} , P^A_{ij} = \braket{p_i p_j}$ and 
$R^A_{ij} = \mathrm{Re}\braket{q_i p_j }$, where the superscript $A$ means that the 
indices $i,j$ are restricted to subsystem $A$. Crucially, a similar strategy can be 
used to construct the R\'enyi negativities (for the details we refer to Ref.~\cite{actc-14}). 
The main idea is that the net effect of the partial transposition with respect to a subinterval 
$A_1$ is the inversion of the signs of the momenta
corresponding to the sites belonging to $A_1$. 

\begin{figure}[t]
\centering
\subfigure
{\includegraphics[width=0.48\textwidth]{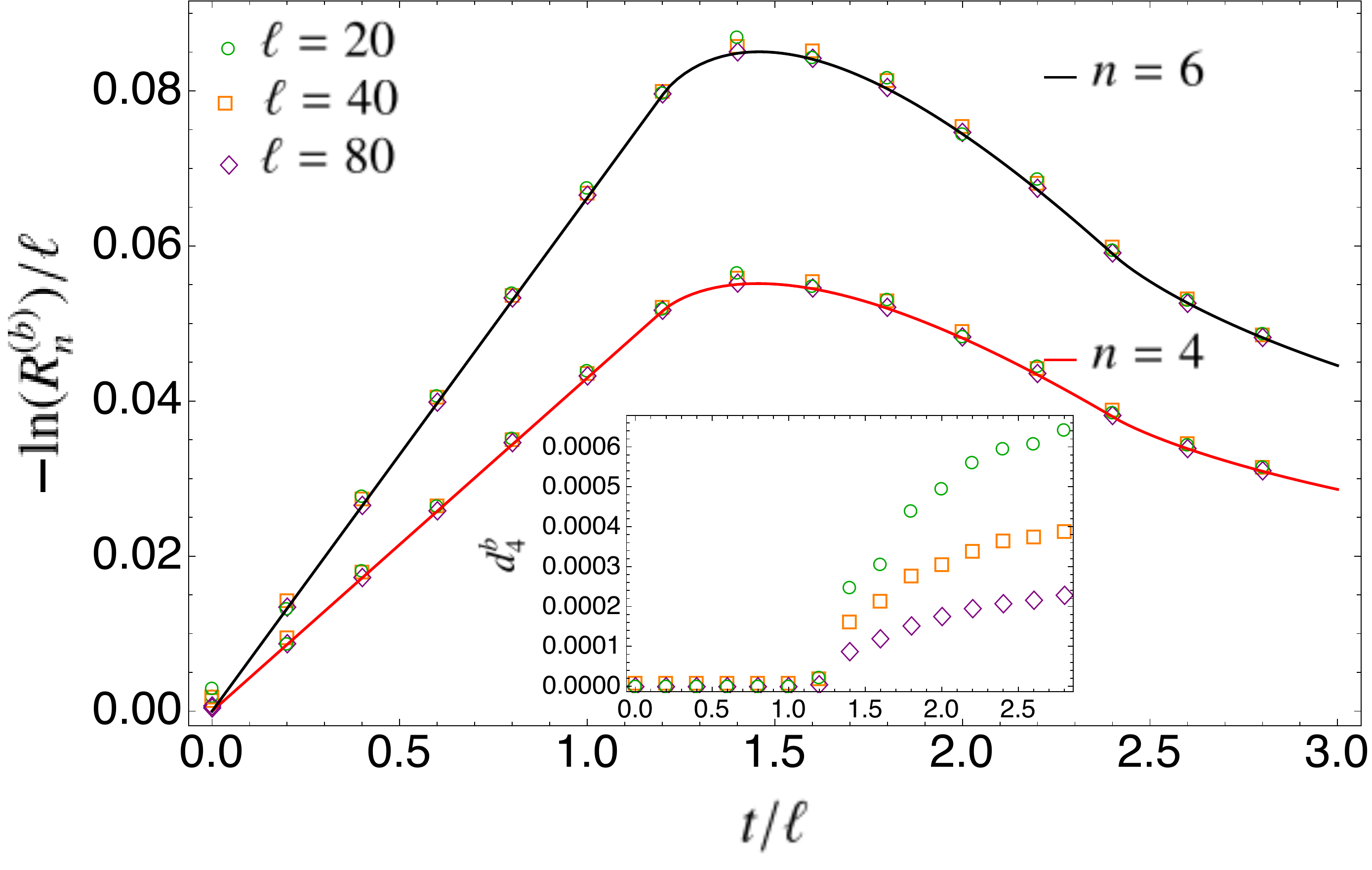}}
\subfigure
{\includegraphics[width=0.48\textwidth]{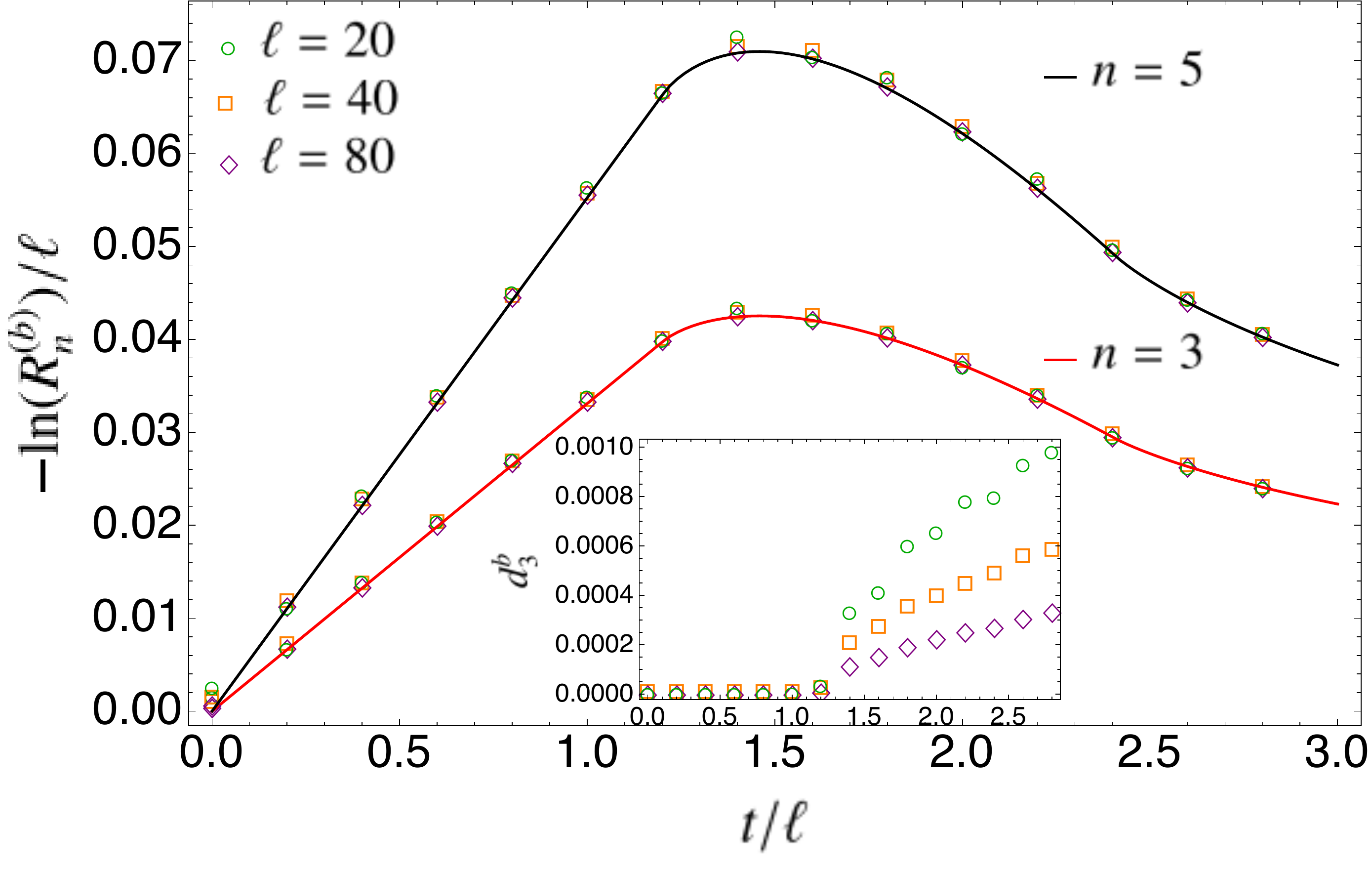}}
\caption{Logarithms of the moments of the (bosonic) partial transpose after the mass quench from $m_0 = 1$ to $m = 2$ in the harmonic chain. 
The quantity $-\mathrm{ln}(R^b_n)/\ell$ is plotted versus rescaled time $t/\ell$, with $\ell$ the intervals' length. 
The analytical predictions represented by continuous lines correspond to Eq.~\eqref{eq:rn}.  
The insets represent Eq.~\eqref{eq:diff} and they prove the validity of  Eq.~\eqref{eq:pure1}, i.e. the connection between the ratio $R_n$ and the R\'enyi mutual information.}
\label{fig:bn}
\end{figure}

For the following, we restrict ourselves to the physical situation with $A$ made of two disjoint parts, 
i.e., $A=A_1\cup A_2$, with  $A_1,A_2$  two equal-length intervals of length $\ell$. We denote as $d$ the 
distance between $A_1$ and $A_2$ (see Fig.~\ref{fig:qp-0}). We only discuss the ratios $-\ln(R_n^{(b)})$ (cf. Eq.~\eqref{eq:ratioRn}). 
The results are shown in Fig. \ref{fig:bn}. Panels (a) and (b) show the quantities $-\ln(R^{(b)}_n)/\ell$ 
for adjacent  intervals, i.e., $d=0$. The data are for several values of the intervals'
length $\ell$ up to $\ell \leq 80$. Since we are interested in the scaling limit, we plot 
$-\mathrm{ln} (R^{(b)}_n)/\ell$ versus the rescaled time $t/\ell$. For two adjacent intervals, the ratio exhibits 
a linear growth for $t/\ell \sim 1.25$, which reflects the maximum velocity being
$v_{\mathrm{max}} \sim 0.4$. For larger times we observe a slow decrease toward
zero for $t/\ell \to \infty$. This slow decay is due to the slower quasiparticles 
with $v < v_{\mathrm{max}}$. The solid line is the theoretical prediction in Eq.~\eqref{eq:rn}. 
At finite $\ell$ and $t$ the data exhibit some small corrections from Eq.~\eqref{eq:rn}, which is recovered 
in the scaling limit $t,\ell\to\infty$ with their ratio fixed. 

It is also useful to investigate directly the validity of Eq.~\eqref{eq:pure1}, which establishes a relationship
between $R_n^{(f/b)}$ and the mutual information. To this aim, we introduce the difference 
$d^{(f/b)}_n$  as 
\begin{equation}\label{eq:diff}
d^{(f/b)}_n=\begin{cases}
\ln(R^{(f/b)}_n) -(1-n)I^{(n/2)}_{A_1:A_2} \quad n\mbox{ even}\\
\ln(R^{(f/b)}_n) -(1-n)\frac{I^{(n)}_{A_1:A_2}}{2},\quad n\mbox{ odd}
\end{cases}
\end{equation}
As it is clear from the insets in Fig.~\ref{fig:bn}, $d_n^{(b)}$ is very small in the region of linear growth, i.e., for  $2v_\mathrm{max} t/\ell\le 1$
(an obvious fact, since in the scaling limit it is just $0-0$). 
At fixed $\ell$, in the non-trivial region, i.e. for larger values of the scaling variable $2v_\mathrm{max} t/\ell $, $d_n^{(b)}$ is larger. 
However, at fixed $t/\ell$, the deviations $d_n^{(b)}$ decrease with increasing $\ell$, and in the scaling limit $\ell\to\infty$ one recovers Eq.~\eqref{eq:pure1}. 
Precisely, the data suggest a behavior $d_n^{(b)} \propto 1/\ell$. 

\subsection{Quench in a free fermion chain}
\label{sec:ising}
We now discuss numerical results for free-fermion systems described by the Hamiltonian
\begin{equation}
H=\sum_{j=1}^{L}\left(\frac{1}{2}[c_j^{\dagger}c^{\dagger}_{j+1}+c_{j+1}c_{j}+c_j^{\dagger}c_{j+1}+c_{j+1}^{\dagger}c_{j}]-hc_{j}^{\dagger}c_{j}\right),
\end{equation}
where $\{c_i,c^{\dagger}_j\}=\delta_{ij}$ are anti-commuting fermionic operators, $h$ is a coupling parameter, e.g. a magnetic field, and we neglect boundary terms (we are interested in the thermodynamic limit $L \to \infty$). 
A Jordan-Wigner transformation maps the Hamiltonian to the well-known transverse field Ising chain. 
However, the spin RDM is not simply mapped to the fermion RDM for two disjoint intervals \cite{ip-09,fc-10}.
Instead, for the case of adjacent intervals they are mapped into each other and so the following results for fermions apply also to the spin variables. 

\begin{figure}[t]
\centering
\subfigure
{\includegraphics[width=0.48\textwidth]{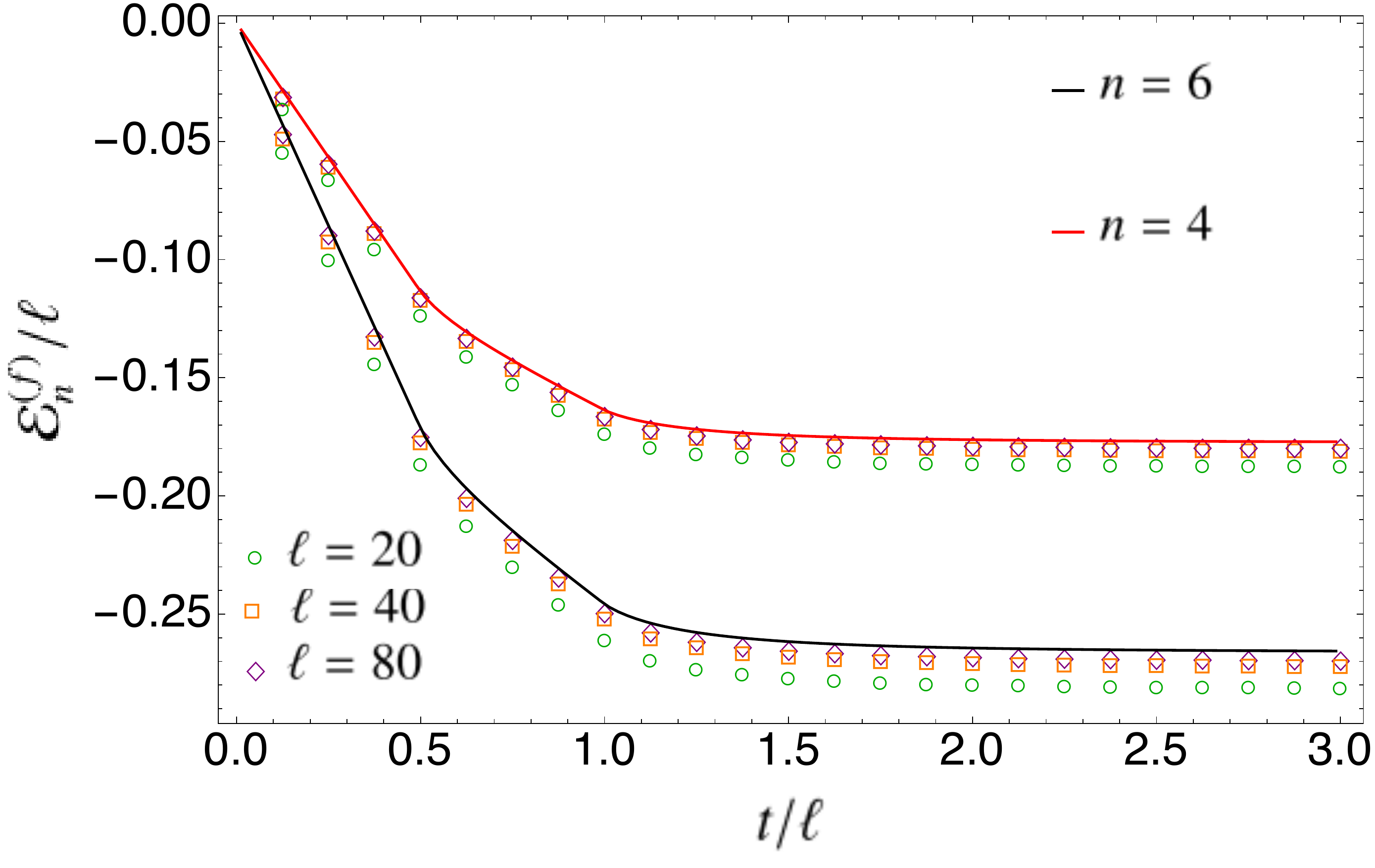}}
\subfigure
{\includegraphics[width=0.48\textwidth]{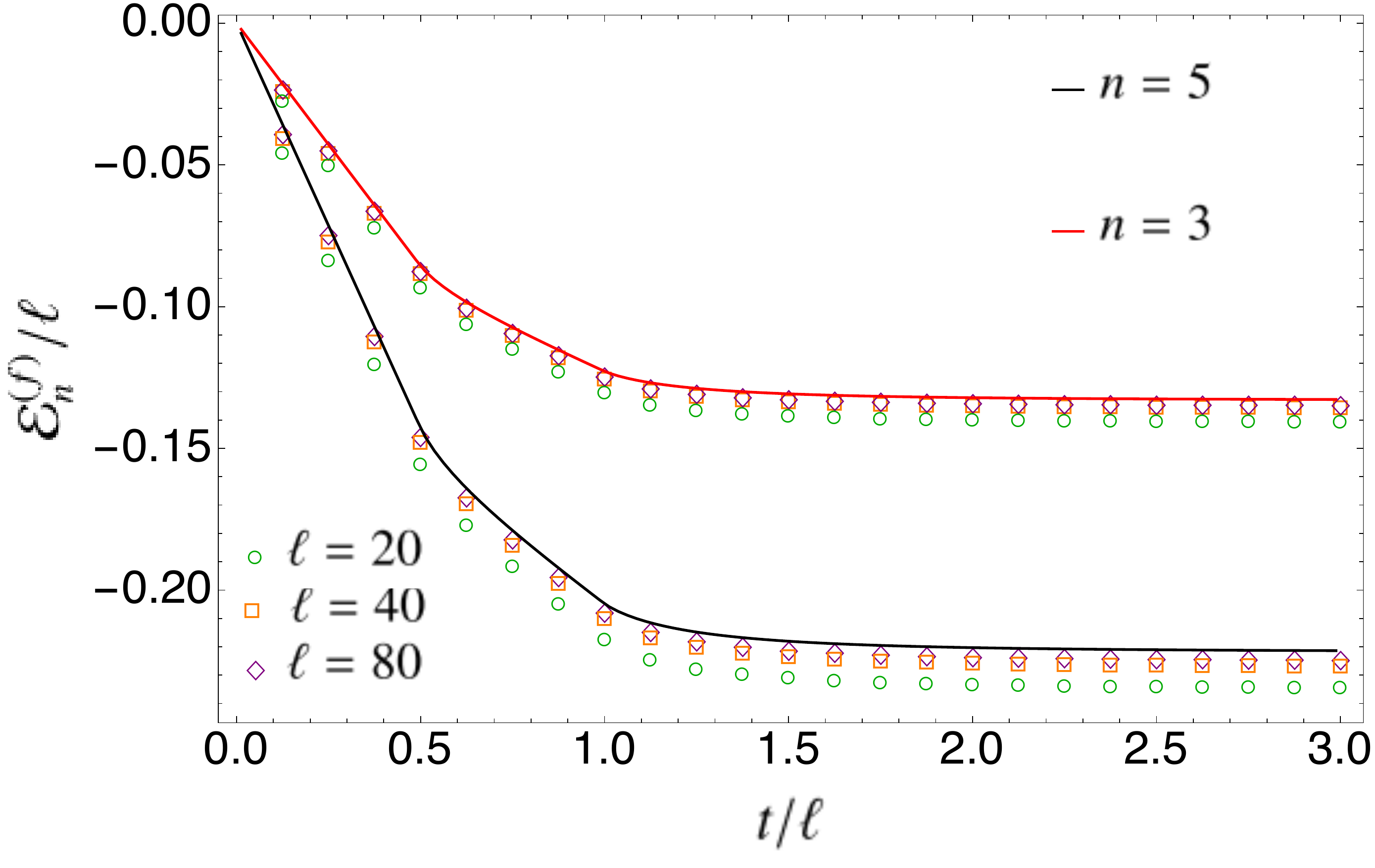}}
\subfigure
{\includegraphics[width=0.48\textwidth]{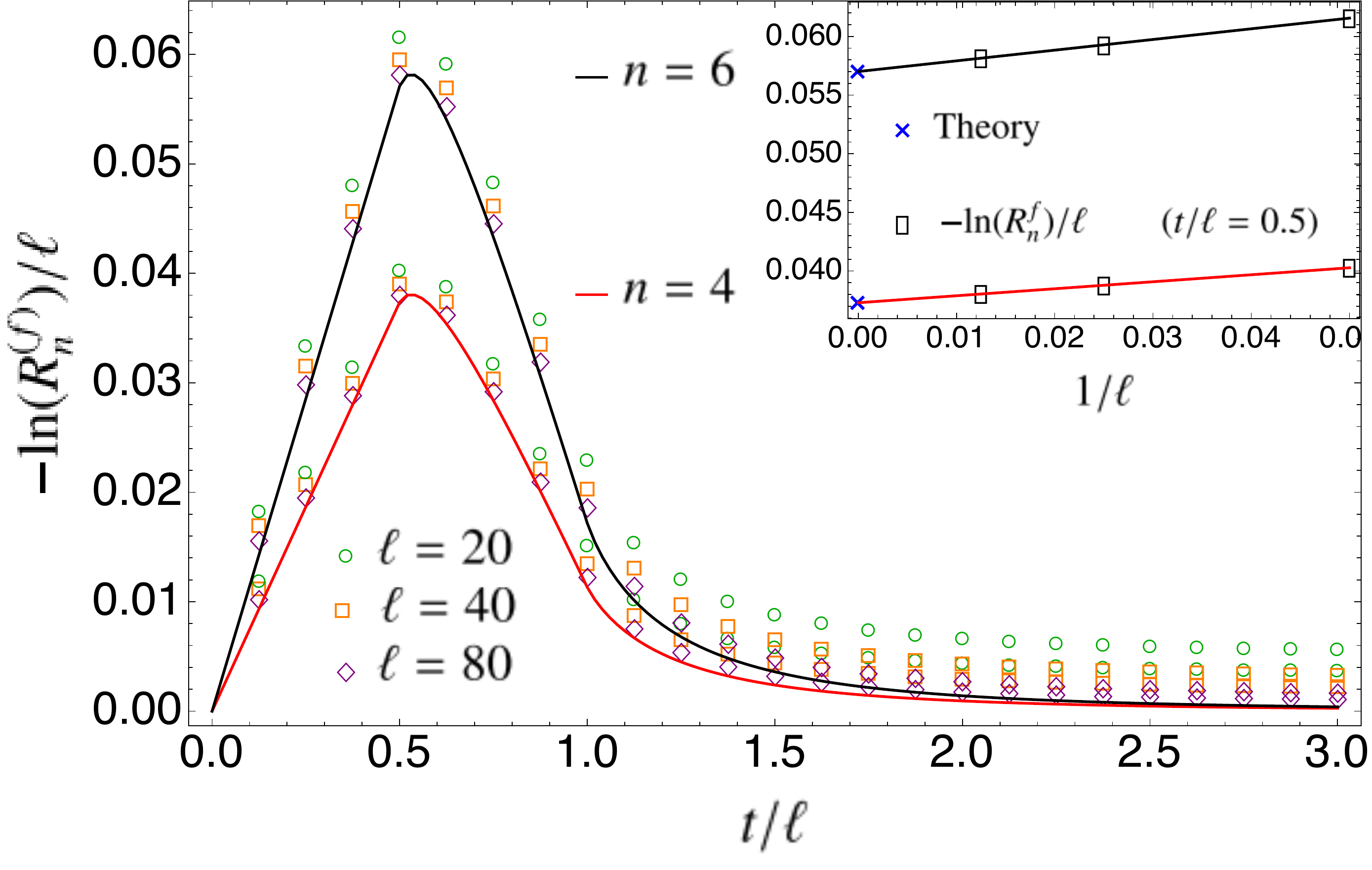}}
\subfigure
{\includegraphics[width=0.48\textwidth]{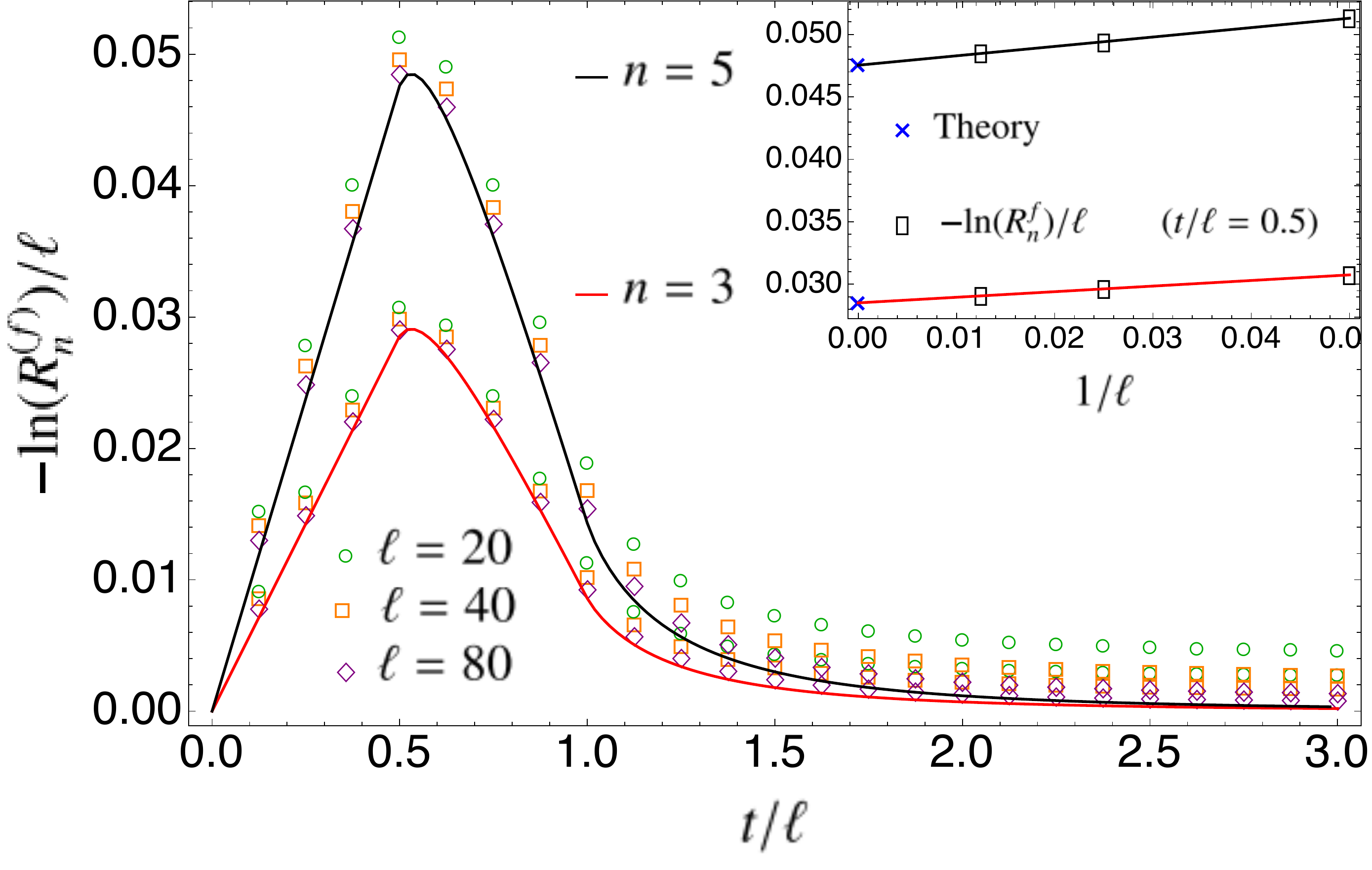}}
\caption{Logarithms of the moments of the (fermionic) partial transpose after a quench in the fermionic chain (with $h_0 = 10$ and $h = 2$) for two adjacent intervals. 
Both $-\mathrm{ln}(R^{(f)}_n)/\ell$ and $\mathcal{E}^{(f)}_n/\ell$ are plotted versus rescaled time $t/\ell$, with $\ell$ the intervals' length. 
The analytical predictions represented by continuous lines correspond to Eq.~\eqref{eq:enwd} (top panels) and \eqref{eq:rn} (bottom panels).  
The insets investigate the finite-size scaling corrections: 
the symbols are for the fermionic negativity at fixed $t/\ell= 0.5 $; the crosses are the theoretical results in the thermodynamic limit; the solid lines are linear fits.}
\label{fig:fn}
\end{figure}

In terms of the momentum space Bogoliubov fermions the Hamiltonian is 
diagonal and the single-particle energies are 
\begin{equation}
e(k) = [h^2 - 2h \cos (k) + 1]^{1/2}.
\end{equation} 
We consider the non-equilibrium unitary dynamics that follows 
from a quench of the field $h$ at $t = 0$ from $h_0$ to $h\neq h_0$. 
In order to parametrise the quench it is useful to introduce the angle $\Delta(k)$ as~\cite{cc-07}
\begin{equation}
	\label{eq:b-angle}
\cos(\Delta(k))=\frac{1+h h_0-(h+h_0)\cos(k)}{e(k)e_0(k)}.
\end{equation}
As for free-bosons, the central object to obtain the quasiparticle prediction is the 
density $\rho(k)$ of the Bogoliubov fermions. This is given by~\cite{cef-11,cef-12}
\begin{equation}
\rho(k)=\frac{1}{2}(1-\cos(\Delta(k))).
\end{equation}
The reduced density matrix can be completely characterized~\cite{ep-09} by the two-point 
correlation functions restricted to the subsystem $A$. From the covariance matrix associated 
to $\rho_A$, one can build the covariance matrix corresponding to the partial time reversal $\rho_A^{R_1}$ (see Ref.~\cite{ssr-17,ssr1-17,ssgr-18,sr-19,ryu,paola}).  
The fermionic R\'enyi negativities $\mathcal{E}^{(f)}_n$ introduced in Eq.~\eqref{eq:sup} 
can be efficiently computed in terms of the eigenvalues of the covariance matrix.

We discuss the numerical results for both $\mathcal{E}^{(f)}_n$ and $R_n^{(f)}$  for two adjacent 
intervals in Fig. \ref{fig:fn}, for the quench with $h_0 = 10$ and $h = 2$. We first discuss the 
R\'enyi negativities $\mathcal{E}^{(f)}_n$ in Fig.~\ref{fig:bn} (top) plotting $\mathcal{E}_n^{(f)}/\ell$ 
versus $t/\ell$. We consider only the geometry with two adjacent intervals. The numerical data exhibit the 
expected behavior as in Fig.~\eqref{fig:qp} (a). One has $\mathcal{E}_n^{(f)}\le 0$ at all times. The negativies 
exhibit a two-slope decrease at short times, which is followed by a saturation at long times. In contrast with the 
CFT case, the saturation is not abrupt due to the fact that the quasiparticles have a nontrivial dispersion. The different 
symbols in the figure denote different subsystem size $\ell$. In the thermodynamic limit $\ell\to\infty$ the 
numerical data approach the prediction of the quasiparticle picture (continuous line in the figure).
We discuss the behavior of the ratios $-\ln(R_n^{(f)})$ in Fig.~\ref{fig:bn} (bottom). Again, we consider only 
the case of adjacent intervals. 
The data for $-\mathrm{ln} (R_n^{(f)})$ exhibit a linear behavior up to $t/\ell\sim 0.5$, reflecting that 
$v_{\mathrm{max}} \sim 1$. Similar to the bosonic case, finite-size corrections are present, 
although the analytical prediction in Eq.~\eqref{eq:quasi2} is recovered in the scaling limit. 
In the inset, we also investigate these scaling corrections. The symbols are the data at fixed 
$t/\ell = 0.5$ while the $x$-axis shows $1/\ell$. The crosses are the theoretical results in 
the scaling limit. The solid lines are fits to the behavior $1/\ell$, 
and are clearly consistent with the data.

\begin{figure}[t]
\centering
{\includegraphics[width=0.55\textwidth]{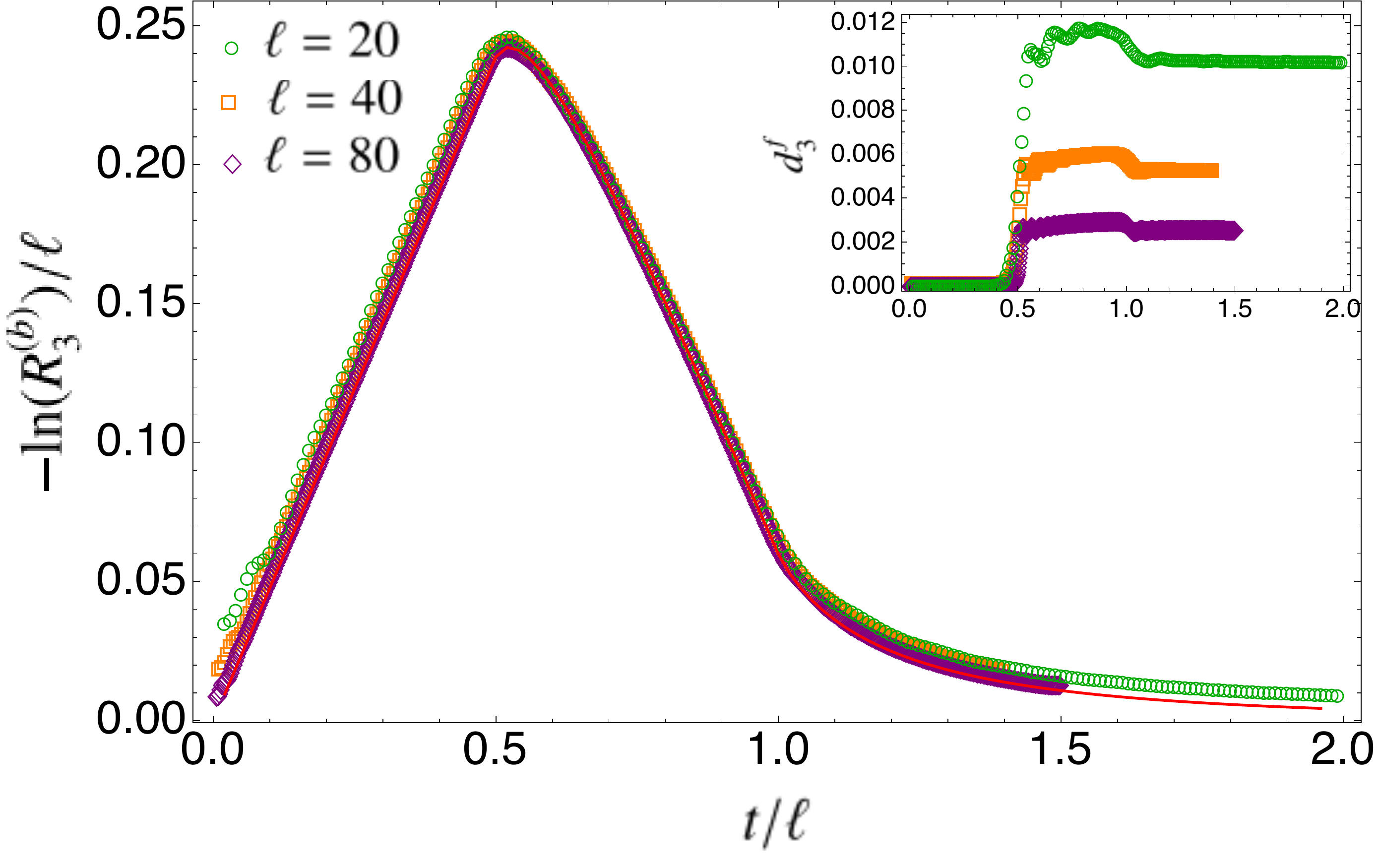}}
\caption{Standard R\'enyi negativity for $n=3$ after a quench in the free fermion chain (with $h_0 = 0.1$ and $h = 2$). 
The numerical data (symbols) have been obtained by expanding the third power of $\rho^{T_1}_A$ as in Eq.~\eqref{eq:rho3} while the solid red line correspond to Eq.~\eqref{eq:rn}. 
The inset represent Eq.~\eqref{eq:diff} and it proves the validity of  Eq.~\eqref{eq:pure1} in the space-time scaling limit. }
\label{fig:fn1}
\end{figure}

We now remind that for free fermion systems (see section~\ref{sec:entmeasure}) it is not straightforward to 
compute the standard negativity defined from the partial transpose $\rho_A^{T_1}$. On the other hand, the 
R\'enyi negativities can be computed effectively. The reason is that the powers of $\rho_A^{T_1}$ can be 
written as sum of products of gaussian operators (cf. Eq.~\eqref{eq:rhoT2}). Since each term of the sum is a gaussian 
operator, its trace can be effectively computed~\cite{fc-10,ez-17}. For instance, for the R\'enyi negativity $\mathcal{E}^{(b)}_3$, we 
have to compute 
\begin{equation}\label{eq:rho3}
\mathrm{Tr}(\rho_A^{T_1})^3=-\frac{1}{2}\mathrm{Tr}(O_+^3)+\frac{3}{2}\mathrm{Tr}(O_+^2O_-).
\end{equation}
By using Eq.~\eqref{eq:rho3} the ratio $-\ln(R^{(b)}_3)$ can be calculated and, for the quench with $h_0=0.1 \to h=2$, 
is reported  in Fig. \ref{fig:fn1}.
The symbols in the figure are numerical results for $-\ln(R^{(b)}_3)$ for two adjacent intervals of length $\ell$. 
The line is the quasiparticle prediction in Eq.~\eqref{eq:quasi2}. The good agreement between the data and the 
analytic curve confirms that $-\ln(R^{(b)}_n)$ and $-\ln(R_n^{(f)})$ are the same in the space-time scaling limit. 
Interestingly, we observe that the fermionic R\'enyi negativity $\mathcal{E}_3^{(f)}$ 
corresponds to the term $\mathrm{Tr}(O_+^2O_-)$ in Eq.~\eqref{eq:rho3}, as can be read in Eq.~\eqref{eq:sup}. 
The fact that the quasiparticle description correctly describes both $\mathrm{Tr}(O_+^2O_-)$ 
and the weighted sum in Eq.~\eqref{eq:rho3}, suggests that the terms in Eq.~\eqref{eq:rho3} become the same 
in the space-time scaling limit. 

Finally, let us shortly discuss the connection between the reduced density matrices of
fermionic and spin models that are connected by the Jordan-Wigner transformation. Due to the non-locality of
this transformation, the reduced density matrices corresponding to $A_1 \cup A_2$ in a spin chain model and its fermionic counterpart
are usually not equivalent unless $A_1$ and $A_2$ are adjacent intervals \cite{ez-17,ctc-15}. 
The same holds also for the (standard) transposed density matrices, for which the identity in Eq.~\eqref{eq:rhoT2} should be slightly modified to take into account the  Jordan-Wigner
string along the interval of length $d$ connecting the two blocks in spin models.

\subsection{Quasiparticle predition for the $p_n$-PPT  conditions}
\label{sec:stilt}

Using the quasiparticle predictions obtained in the previous sections for the R\'enyi negativity, 
one can write down the quasiparticle formulas for the $p_n$-PPT conditions introduced in 
section~\ref{sec:neg}, see Eq. \eqref{DDn}. 
For instance, the $p_3$-PPT condition quantifies the violation of Eq.~\eqref{eq:p3}. 
Specifically, the condition  $D_3\equiv p_3-p_2^2<0$ signals the presence of 
quantum entanglement. As explained in section~\ref{sec:neg}, other conditions $D_n\geq 0$ can be 
obtained by considering higher moments of the partial transpose. 

\begin{figure}[t]
\centering
\subfigure
{\includegraphics[width=0.47\textwidth]{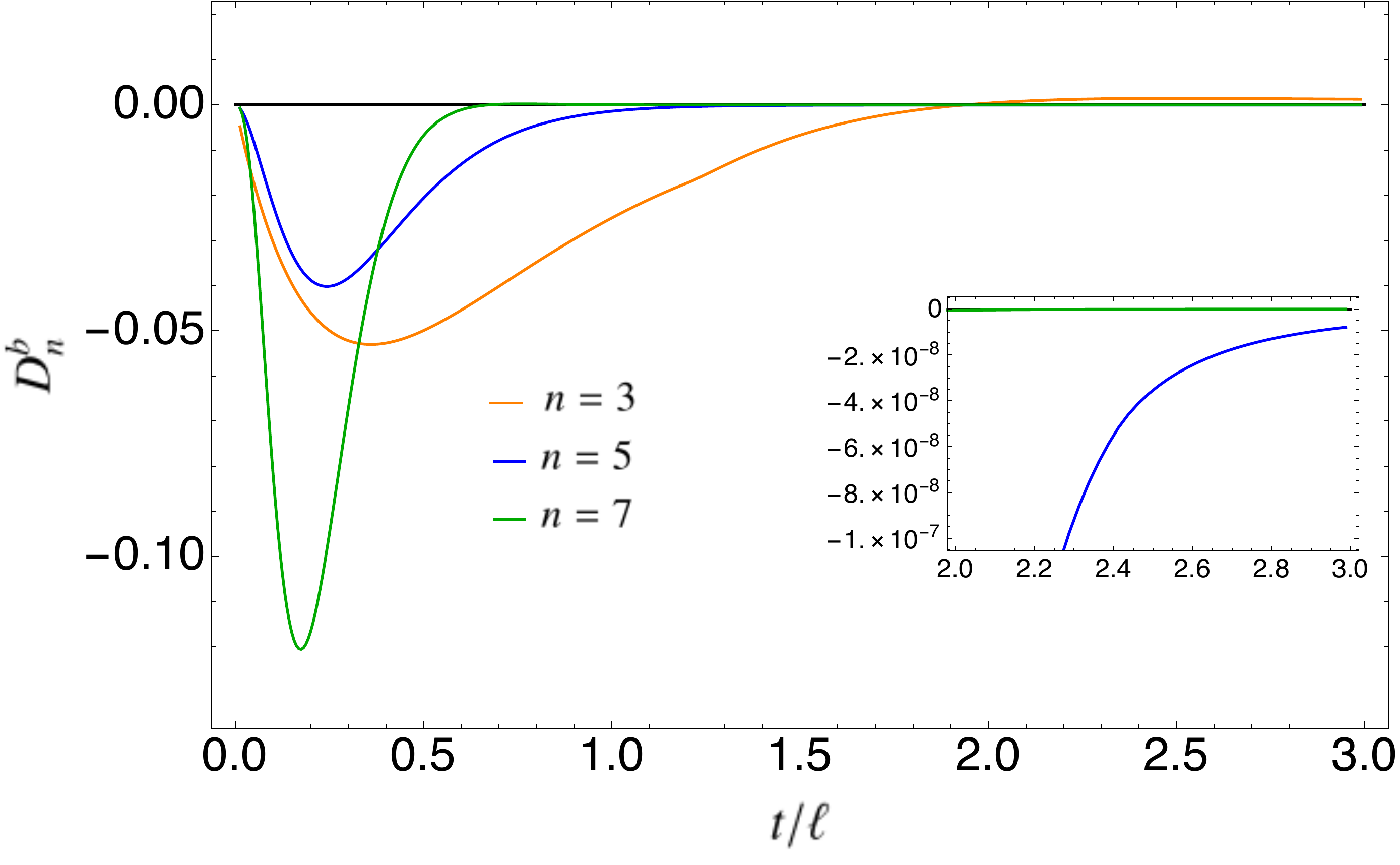}}
\subfigure
{\includegraphics[width=0.48\textwidth]{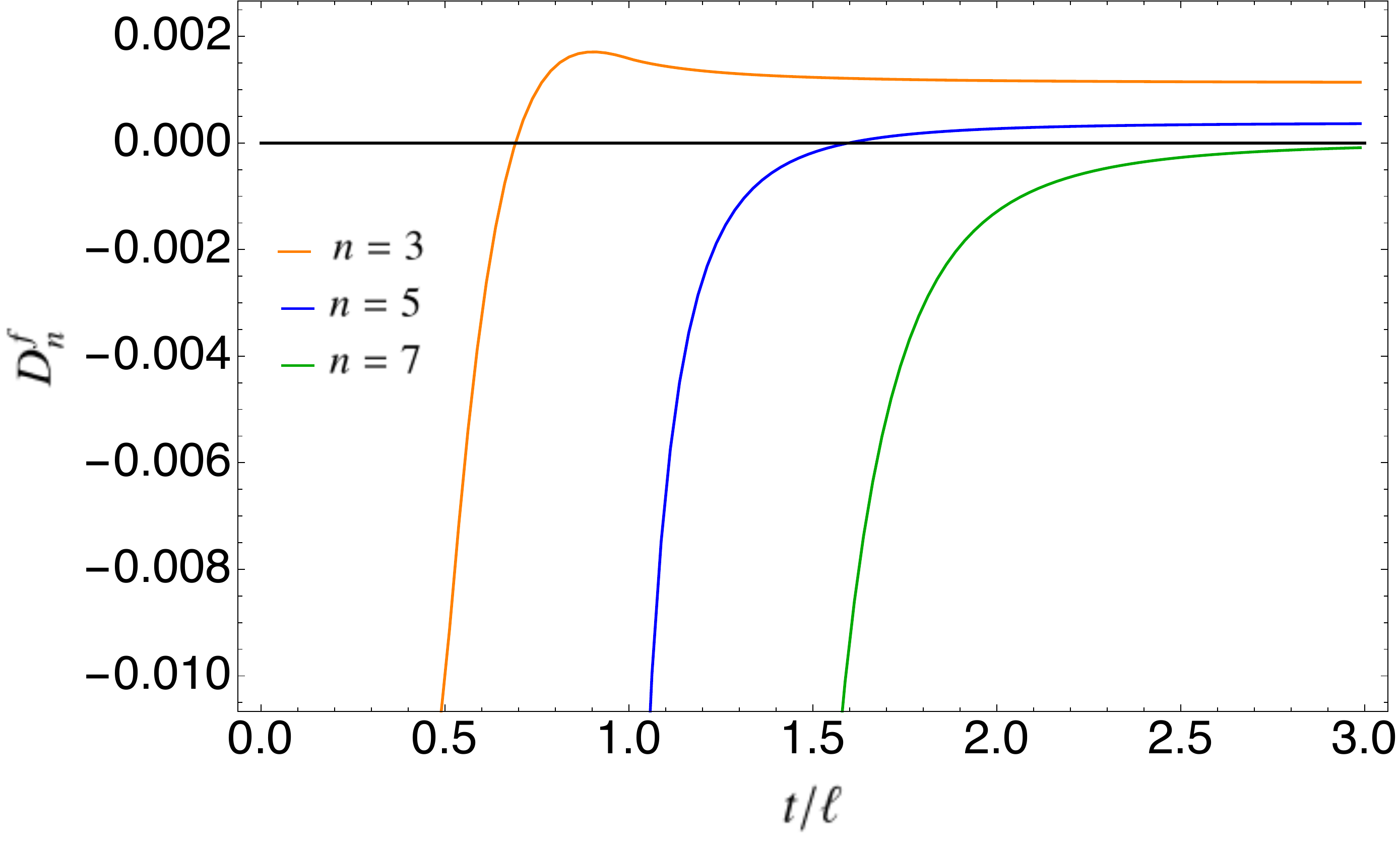}}
\caption{The $p_n$-PPT conditions for $n=3,5,7$ for the quasiparticle predictions of the moments of the partial transpose introduced in Eq.~\eqref{eq:pmoments}. 
The quench parameters are $h_0=10, h=2$ for the fermionic chain and $m=2,m_0=0.1$ for the harmonic chain. 
We plot the quantity $D_n$ in Eq. \eqref{DDn}. The $p_n$-PPT condition is $D_n\geq 0$. 
To compare data for different $n$ we multiply the $D_5$ by 10 and  $D_7$ by $10^6$ for bosons ($\ell=30$).
For fermions the $D_5$ is multiplied by $10^7$ and $D_7$ by $10^{19}$ ($\ell=50$).
The violation of these conditions for at least one value of $n$ reveals the presence of entanglement between $A_1$ and $A_2$.}
\label{fig:criteria}
\end{figure}

We numerically investigate the $p_n$-PPT conditions in Fig.~\ref{fig:criteria} for $n=3,5,7$, 
and for quenches in both the fermionic and  harmonic chains. We focus on the situation with two adjacent intervals and we 
are interested in understanding how the $p_n$-PPT conditions are violated as a function of time. 
The results in Fig.~\ref{fig:criteria} are obtained by using the quasiparticle picture prediction. 
As it is clear from the figure, all $p_n$-PPT conditions are violated at short times in both models. 
At infinite times all the $p_n$-PPT conditions give zero. 
These results are consistent with the behavior of the logarithmic negativity~\cite{ac-18}. 

Also the fine structure of these $p_n$-PPT conditions is very interesting. 
In the short-time region with a lot of entanglement (compare with the previous figures for $R_n$) all the conditions are violated.
As the time increase and the entanglement becomes much less, the first condition to be satisfied is $p_3$ (i.e. $D_3>0$) and only after the other one
(the panel on the left is particularly clear in this respect)
This implies that higher and higher $p_n$-PPT conditions are necessary to detect the very little amount of entanglement present at large time. 
This fact is not surprising, but it is remarkable that it is captured so neatly by the quasiparticle picture.

\section{Conclusions}\label{sec:conclusions}

In this paper, we derived the quasiparticle picture for the dynamics 
of the moments of the partially transposed reduced density matrix  after a
quantum quench in integrable systems, and several related quantities such as 
the R\'enyi negativities $\mathcal{E}_n$ (cf. Eq.~\eqref{eq:renneg}) and the ratios $R_n$ (cf. Eq.~\eqref{eq:ratioRn}). 
An interesting result is that the ratio $R_n$ is proportional to the R\'enyi mutual information. 
Furthermore, this ratio is qualitatively similar to the negativity and so it is then an indicator of the 
entanglement barrier for the quench dynamics at intermediate time \cite{ac-18,d-17,lpb-18,jhn-18}.
Moreover, our results allow us to derive the behavior of the  $p_n$-PPT conditions, which  
in contrast with standard entanglement measures for mixed states, such as the logarithmic negativity, 
are easily computable and experimentally measurable for quantum many-body systems~\cite{ekh-20,ncv-21}. 
We tested our predictions against exact numerical results for both free-fermion and free-boson 
systems. 

We now discuss future research directions. 
The first natural followup of the results presented here is to test numerically the equality between R\'enyi mutual informations and the ratios $R_n$ for interacting integrable models. 
A possible extension to this work would be to study the dynamics of negativity and the moments of the partial transpose in the 
presence of a globally conserved charge. While it is known that the negativity of two subsystems may be
decomposed into contributions associated with their charge imbalance \cite{csg-19,mbc-21}, 
it would be interesting to understand whether a quasiparticle prediction similar to the one presented for the entropy in Ref.\cite{pbc-20} 
could be worked out.
Another important research direction is to investigate the R\'enyi negativities and the ratios $R_n$ in the presence of dissipation~\cite{carollo-d}. 

\section*{Acknowledgments}
P.C. and S.M. acknowledge support from ERC under Consolidator grant number 771536 (NEMO).  
V.A. acknowledges support from the European Research Council under ERC Advanced grant No. 743032 DYNAMINT.


\end{document}